\documentclass{aa}

\usepackage{amsmath, amssymb, amsfonts}
\usepackage{graphicx}
\usepackage{natbib}
\usepackage{hyperref}

\usepackage[caption=false]{subfig}
\usepackage{multirow}

\usepackage[T1]{fontenc}
\usepackage{cleveref}
\usepackage{siunitx}
\usepackage{tabularx}
\usepackage[dvipsnames]{xcolor}
\usepackage{float}
\usepackage{bm}
\usepackage{gensymb}
\usepackage[normalem]{ulem}
\usepackage{txfonts}
\usepackage{diagbox}
\usepackage{soul}
\usepackage{comment}
\crefname{section}{§}{§§}
\Crefname{section}{§}{§§}
\crefformat{section}{§#2#1#3}

\begin{document}
        
        \title{Comparing radial migration in dark matter and MOND regimes}
        
        \subtitle{}
        
        \author{R. Nagy \inst{1}\thanks{E-mail: roman.nagy@fmph.uniba.sk} \and F. Jan\'ak\inst{1} \and M. \v{S}turc\inst{1}      \and M. Jur\v{c}\'ik\inst{1,2} \and E. Puha \inst{1}}
        
        \authorrunning{Nagy et al.}
        
        \institute{
                Faculty of Mathematics, Physics, and Informatics, Comenius University, Mlynsk\'{a} dolina, 842 48 Bratislava, Slovakia
        \and
                Astros Solutions s. r. o.
        Nad L\'{u}\v{c}kami 35, 841 04 Bratislava, Slovakia}
        
        \date{Received xxxx; accepted xxxx}
        
        \abstract
        {Multiple studies on radial migration in disc galaxies have proven the importance of the effect of resonances with non-axisymmetric components on the evolution of galactic discs. However, the dynamical effects of classic Newtonian dynamics with dark matter (DM) differ from MOdified Newtonian Dynamics (MOND) and might trigger different radial migration. A thorough analysis of radial migration considering these two gravitational regimes might shed some light on different predictions of DM and MOND theories.}
        {We aim to quantitatively and qualitatively compare the effects of resonances and stellar radial migration (churning) in a Milky Way-like (MW-like) galaxy in the DM and MOND regimes. We performed numerical simulations of a MW-like galaxy to analyse the effect of non-axisymmetric structures (galactic bar and spiral arms) considering various parameters of the spiral structure.}
        {We conducted a two-dimensional numerical simulation consisting of the integration of $2 \cdot 10^6$ stars in a static rotating galactic potential for $6~\mbox{Gyr}$. We analysed the changes in the stellar disc caused by stellar radial migration. To study the effects of radial migration and resonances in detail, we analysed the change in the star's position, the guiding radius, as well as the frequency phase space. We investigated DM and MOND approaches.}
        {The outcome of the simulation shows that the radial migration is much more pronounced in the MOND regime compared to the DM one. Increasing both the spiral amplitude and the spiral pattern speed results in more prominent radial migration.
        Compared to the DM approach, in the MOND regime, we observe up to five times as many stars with a maximum change in the guiding radius of more than $1.5~\mbox{kpc}$ during the time interval from $2-6~\mbox{Gyr}$. Increasing the amplitude of the spiral structure amplifies this effect. On the other hand, increasing the spiral pattern speed reduces the difference between DM and MOND.
        Analysis of the frequency phase space reveals that the most prominent resonances in all DM and MOND configurations are the co-rotation resonance with the spiral arms ($m=p=1$), outer Lindblad resonance with the galactic bar and spiral arms, and the co-rotation resonance ($m=2$, $p=1$) with the superposition of the galactic bar and spiral arms, $2 \Omega = \Omega_b + \Omega_{sp}$.}
        {}
        \maketitle

\section{Introduction}
Radial migration plays a crucial role in the evolution of the galactic discs, as it can substantially displace stars from their birth radius. It is described as a change in the angular momentum of an individual star (a change in the guiding radius, i.e. churning) \citep{Schoenrich2009a,Schoenrich2009b}, while the overall distribution of angular momentum of the disc does not change significantly \citep{sellwood2002radial}. Moreover, stellar orbits can be altered by blurring, which is characterised by an increase in the star's epicycle amplitude and eccentricity \citep{Schoenrich2009a,Schoenrich2009b}.

Models and simulations show that radial migration can occur due to external effects, such as perturbation from galactic mergers \citep{2009MNRAS.397.1599Q, Donlon2024, Merrow2024} or under the effect of non-axisymmetric features. Resonances with  spiral arms \citep[e.g.][]{sellwood2002radial, Roskar2008b, vera-ciro2014, 2015MNRAS.447.3576D, Bautista2021} and the galactic bar \citep[e.g.][]{Ceverino2007, 2011A&A...534A..75B, 2013A&A...553A.102D, 2013MNRAS.436.1479K, Trick2021} are especially powerful in inducing significant radial migration.
The combined effect of both the galactic bar and the spiral arms was explored as well \citep{minchev2010, Minchev2011, halle2018, 2019ApJ...882..111D, Feltzing2020}. \cite{minchev2010} showed that resonances of such overlap can trigger a strong non-linear response (the change in orbital angular momentum at the resonance overlap is larger than the sum from each resonance) that significantly redistributes the angular momentum of the disc. The contribution of the spiral arms to overall radial migration appears to be more important than that of the bar. 
\cite{klacka2012} reported that the influence of the spiral structure on the radial migration of the Sun is much more prominent than the effect of the Galactic bar, which is negligible. Radial migration is also affected by specific parameters of the non-axisymmetric structures. For instance, \citet{Okalidis2022} studied 17 simulated Milky Way-like (MW-like) galaxies and found that galaxies with a strong bar show larger radial migration. \citet{Bautista2021} focussed on the spiral arms, exploring how the different spiral parameters influence radial migration. They found that the scale length of the spiral arms is the most relevant parameter of the spiral structure that affects radial migration. Interaction with tight-wound spiral arms can lead to the formation of regions of higher stellar concentrations for stars co-rotating with the spiral arms \citep{quillen-minchev2005} and might trigger a large radial migration in the resonance zones \citep{Trick2022}. Simulating a barred spiral galaxy, \cite{2015A&A...578A..58H} found significant churning caused by the co-rotation resonance of the bar. The authors also report that the outer Lindblad resonance (OLR) separates the disc into two distinct regions with minimal exchange of angular momentum. The simulations performed by \cite{halle2018} discovered that both the thin and thick disc experience a similar amplitude of churning and that their stars can be trapped in a co-rotation resonance with the bar.

Radial migration has a profound impact on the galactic disc and is necessary to explain phenomena such as the metallicity dependence on the Galactocentric radius \citep[e.g.][]{Schoenrich2009a,Loebmann2016,Anders2017,Zhang2021,Okalidis2022}, the upturn of the stellar age gradient on the outskirts of local galaxies \citep{Bakos2008,Roskar2008b,Yoachim2012,Riuz-Lara2017}, or the scatter in the age-metallicity distribution \citep[e.g.][]{Haywood_2008,Roskar2008,Wang2013,Kubryk2015,Lu2022}. Moreover, radial migration was proposed as the mechanism that forms the thick disc \citep[e.g.][]{Schoenrich2009a,Schoenrich2009b,Loebman2011,Roskar2013,Kubryk2015}, although there are also contradictory results showing that radial migration is ineffective in disc thickening \citep[e.g.][]{Minchev2012,Minchev2014,Grand2016b}. Considering the impact of radial migration on the vertical structure of the galactic disc, \cite{Solway2012} found that stars more likely to migrate are close to the midplane and \cite{vera-ciro2014} showed that radial migration is heavily biased towards stars with low vertical velocity dispersions.

Although the impact of radial migration on the disc is significant, it is difficult to quantify it exactly. \cite{Frankel2018} show that radial migration in the Milky Way (MW) disc is strong, with stars typically migrating by the half-mass radius of the MW disc over the age of the disc. \cite{Feltzing2020} use data from spectroscopic surveys to show that $50-60\%$ of stars have radially migrated, while around 10\% of stars experienced only churning. Further, they report that only less than 7\% of stars have never undergone either churning or blurring. \cite{Frankel20} found that the evolution of the MW is dominated by diffusion in angular momentum, while blurring is an order of magnitude lower. \cite{Lian22} found that around half of MW stars at ages of 2 and 3 Gyr have migrated more than 1 kpc from their formation radius at ages of 2 and 3 Gyr, although only less than 5\% of stars have migrated more than 4 kpc. By analyzing metal-rich red giants from APOGEE, \cite{2024MNRAS.533..538L} found that the majority of metal-rich stars have experienced churning, while about half of super-metal-rich stars have experienced blurring. Several papers have quantified radial migration in the MW based on open clusters. \cite{Chen20} found that 56\% of open clusters have experienced significant churning, while the younger clusters tend to migrate inwards and older clusters outwards, which is also supported by the results of \cite{Zhang2021}. \cite{2023A&A...679A.122V} investigated the difference in radial migration of MW open clusters and MW field stars. They report that older clusters have more inclined and less circular orbits than isolated stars, while younger clusters are more resistant to perturbations than field stars of the same age.

So far, radial migration has been explored only in Newtonian gravity. However, in alternative theories, such as MOND, radial migration might look different. Therefore, in this work, we explore how the classic Newtonian approach and MOND affect radial migration. We make a simplified simulation of a MW-like galaxy, where we can clearly observe the differences between the two approaches. We simulate a MW-like galaxy as a static two-dimensional potential using the classic dark matter (DM) approach with the Navarro-Frenk-White (NFW) halo. On the other hand, we also consider MOdified Newtonian Dynamics (MOND), adopting the formula introduced by \citet{mcgaugh2016}. This approach is based on the fit of the rotation curves of spiral galaxies from the Spitzer Photometry \& Accurate Rotation Curves, \citep[SPARC]{SPARCmaster} database, which are similar to the MW galaxy.

The paper is organised as follows. In Sec. \ref{Models}, we present the potentials of the components of the galaxy, the DM halo for the DM approach, and the MOND formula. In Sec. \ref{Simulations}, we introduce our simulation, initial conditions, and numerical methods. In Sec.\ref{Resonances}, we present the theory of resonances focusing on the co-rotation and Lindblad resonances. In Sec.\ref{Results}, we present the results of the simulation and compare the differences between the DM and MOND approaches. In Sec. \ref{Discussion}, we discuss the limitations of our work and compare our results with the published literature. In Sec. \ref{Conclusions}, we conclude this work.

\section{Models}\label{Models}
We modelled the trajectory of every particle in the galactic disc separately. For simplicity, we studied only the two-dimensional motion in the plane. We used the polar galactocentric coordinate system with a galactocentric distance, $R$, and galactic azimuth, $\varphi$, which was measured in the opposite direction as the rotation of the spiral arms and the galactic bar. We considered two approaches, one consisting of the classical Newtonian approach with a DM halo, and the other being the MOND approach. In the DM approach, the equation of motion can be written as

\begin{equation}\label{Newt}
\dot{\vec{v}} \,=\, - \nabla (\Phi_{b}+\Phi_{d}+\Phi_{sp}+\Phi_{h})\,,
\end{equation} 
where we decomposed the galactic potential in four components - the bulge, $\Phi_{b}$ (including the bar), the disc, $\Phi_{d}$, the spiral arms, $\Phi_{sp}$, and the DM halo, $\Phi_{h}$.

In the case of MOND, a modification of the equation of motion was necessary. We adopted the approach of \cite{mcgaugh2016} in the following vector generalisation \citep[e.g.][]{Klacka2019}:
\begin{equation}\label{nonNewt}
\begin{aligned}
\dot{\vec{v}} \,&=\,\frac{\vec{g}_{bar}}{1-\mbox{exp}(-\sqrt{g_{bar}/g_{\dag}})}\,, \\[.5cm]
\vec{g}_{bar} \,&=\, -\nabla(\Phi_{b}+\Phi_{d}+\Phi_{sp})\,,
\end{aligned}
\end{equation}
where $g_{\dag}=1.2\cdot10^{-10}~\mbox{m}~\mbox{s}^{-2}$, ${\vec{g}}_{bar}$ is the acceleration due to the baryonic matter and the components are the same as in the Newtonian case, except for the halo, which is not present in MOND. The potentials of each component are described below.

\begin{table}
\caption{\centering Simulation parameters.}
        \label{parameters}
        \centering
        {\small
                \begin{tabular}{|c|c|c|c|}\hline
                        Structure & Parameter & Value & Units\\ \hline\hline
   Bulge and bar & $M_{b}$ & $1.4\cdot10^{10}$ & M$_{\odot}$ \\
   & $\Omega_{b}$ & $55.5$ & $\mbox{km}~\mbox{s}^{-1}~\mbox{kpc}^{-1}$ \\
   & $r_{b}$ & $3.44$ & $\mbox{kpc}$ \\
   & $Q_{11}$ & $1.86\cdot M_{b}$ & $\mbox{M}_\odot~\mbox{kpc}^{2}$\\
   & $Q_{22}/Q_{11}$ & $-0.429$ & ---\\ \hline
   
   Disc & $\Sigma_0$ & $896$ & {M}$_{\odot}~\mbox{pc}^{-2}$ \\ 
   & $r_d$ & $2.50$ & $\mbox{kpc}$ \\ \hline
   
   Spiral arms & $N$ & $100$ & ---\\ 
   & $i$ (locus1) & $12.8$ & $\mbox{deg}$ \\
   & $i$ (locus2) & $15.5$ & $\mbox{deg}$ \\
   & $r_{sp}$ (locus1) & $3.6$ & $\mbox{kpc}$ \\
   & $r_{sp}$ (locus2) & $2.6$ & $\mbox{kpc}$ \\\hline
   
   DM halo & $\rho_{0,h}$ & $8.54 \cdot 10^{-3}$ & $\mbox{M}_{\odot}~\mbox{pc}^{-3}$ \\
   & $r_{h}$ & $19.6$ & $\mbox{kpc}$ \\\hline   
                \end{tabular}   }
\tablefoot{Values of the parameters of galactic components used in all simulations.}
\end{table}

\begin{table}
\caption{\centering Simulation names.}
\label{parameters2}

\resizebox{\columnwidth}{!}{
        \centering
                \begin{tabular}{|l|l|c|c|}\hline
                        simulation label & regime & $\Omega_{sp}$ [$\mbox{km}~\mbox{s}^{-1}~\mbox{kpc}^{-1}$] & $A_{sp}$ [$\mbox{km}^{2}~\mbox{s}^{-2}~\mbox{kpc}^{-1}$]\\ \hline\hline
   DM\_19\_1000 & Dark Matter & 19 & 1000 \\
   DM\_25\_850 & Dark Matter & 25 & 850 \\
   DM\_25\_1000 & Dark Matter & 25 & 1000 \\
   DM\_25\_1200 & Dark Matter & 25 & 1200 \\
   DM\_31\_1000 & Dark Matter & 31 & 1000 \\
   \hline
   MOND\_19\_1000 & MOND & 19 & 1000 \\
   MOND\_25\_850 & MOND & 25 & 850 \\
   MOND\_25\_1000 & MOND & 25 & 1000 \\
   MOND\_25\_1200 & MOND & 25 & 1200 \\
   MOND\_31\_1000 & MOND & 31 & 1000 \\\hline
                \end{tabular}   }
\tablefoot{Names of the simulation runs with sets of $\Omega_{sp}$ and $A_{sp}$ parameters used in the simulation run.}
\end{table}

\subsection{Galactic disc}\label{secDisc}
We used the exponential disc model from \citet[Ch. 2.6, 2.164a]{binney-tremaine2008}:

\begin{equation}\label{potedisc}
\Phi_{d}(R) = -\pi G\Sigma_{0}R\left[I_{0}(y)K_{1}(y)-I_{1}(y)K_{0}(y)\right]\,,
\end{equation}
where $y\equiv R/2r_{d}$, $r_{d}$ is the scale length of the disc, $G$ is the gravitation constant, $\Sigma_{0}$ is the planar density of the galactic disc, and $I_{n}$ and $K_{n}$ are modified Bessel functions. The parameters of the disc, adopted from \cite{mcmillan2017}, are given in Table \ref{parameters}. As we studied the trajectories in the plane only, we took into account only the thin disc.

\subsection{Galactic halo}
For the DM halo, we used the commonly used NFW model \citep[]{nfw1996}, for which the potential has the form
 
\begin{equation}\label{poteHalo}
\Phi_{h}(r) = -\frac{4\pi G r_{h}^{3} \rho_{0,h}}{r}\ln\left(1+\frac{r}{r_{h}}\right)\,,
\end{equation}
where $r$ is the spherical galactocentric distance, $\rho_{0,h}$ is the density of the halo, and $r_{h}$ is the scale length. Parameters are adopted from \cite{mcmillan2017} and are given in Table \ref{parameters}.

\subsection{Spiral arms}
For the spiral arms, we used the tight-winding approximation \citep{contopoulos1986}, in which the potential of the spiral pattern has the following form:
\begin{equation}\label{poteSpiral}
\Phi_{sp}(r,\varphi)=-A_{sp}r\,\mbox{e}^{-r/r_{d}}\cos[2(\varphi+\Omega_{sp}t)-g(r)]\,,
\end{equation}
where $t$ is time, $A_{sp}$ is the amplitude of the spiral pattern, $\Omega_{sp}$ is its spiral pattern speed, and the function $g(r)$ defines the spiral shape. The spiral shape defined by \cite{roberts1979} is
\begin{equation}\label{locus}
g(r)=\frac{2}{N\tan(i)}\ln\left[1+\left(\frac{r}{r_{sp}}\right)^{N}\right]\,,
\end{equation}
where $r_{sp}$ is the beginning radius of the pattern and $i$ is the pitch angle. We followed the approach of \cite{antoja2011} and considered a spiral structure with two loci. The parameter $N$ measures how sharply the spiral structure changes to the bar structure in the central regions. The limiting case of $N\rightarrow\infty$ is here approximated by $N=100$, which creates spiral arms that begin to form an angle of $\sim90^{\circ}$ with the line that joins the two starting points of the locus. The values of the parameters adopted from \cite{antoja2011} can be found in Tables \ref{parameters} and \ref{parameters2}.

We investigated two different spiral shapes (see values $r_{sp}$ and $i$ for locuses 1 and 2 in Table \ref{parameters}). However, our simulations have shown that the influence of the spiral shape on the final results is negligible (we did not find any remarkable quantitative or qualitative difference between locuses 1 and 2). Therefore, we decided to continue with the spiral shape described as locus 1 in the Table \ref{parameters}.

\subsection{Central bar}

We used a model of the galactic bar developed by \cite{klacka2012} for the non-axisymmetric central part of the MW, also applicable to our MW-like galaxy. This model consists of a multipole expansion of the gravitational potential generated by a distribution of matter presented by \cite{freudenreich1998}, based on the COBE data. The multipole expansion up to $\Phi^{(3)}$ describes the gravitational potential outside the galactic bulge ($r>r_b$, where $r_b$ is the maximum extent of the bulge). We did not consider the area inside the bulge (see Sec. \ref{unfinished}). Our model has the following form:

\begin{equation}
\begin{aligned}
&\Phi_{b}(r,\varphi)= - \frac{G M_{b}}{r} - \frac{G K}{4 r^{3}},\\
&K = Q_{11} \left( 1 - Q_{22}/Q_{11} \right) \left[ \frac{1 + Q_{22}/Q_{11}}{1 - Q_{22}/Q_{11}} + \cos\left(2\Omega_{b}t+2\varphi\right) \right],
\end{aligned}
\end{equation}
where $M_{b}$ is the mass of the bulge, $Q_{11}$ and $Q_{22}$ describe the non-spherical mass distribution, and $\Omega_{b}$ is the bar pattern speed. The values of the parameters, adopted from \citet{klacka2012}, are given in Table \ref{parameters}.

\section{Numerical simulation}\label{Simulations}

\subsection{Numerical integration}\label{Num-int}

Our simulation represents a two-dimensional motion of stars in an external force field. It consists of numerical integration of equations \ref{Newt} and \ref{nonNewt} for each particle, obtaining a time evolution of the state vector. Our integration lasted $6~\mbox{Gyr}$ with a time step of $2.5~\mbox{Myr}$. In our case, the chosen time step value is sufficient enough (positions and velocities of even the most extreme stars typically differ by less than $1\%$ after forward-backward integration). All calculations were performed in the polar galactocentric coordinate system. We simulated five cases with different values of $\Omega_{sp}$ and $A_{sp}$, each in the DM and MOND configuration. All ten simulations share the parameters listed in Table \ref{parameters}. Table \ref{parameters2} shows the names and the specific parameters of each simulation runs. As we aim to study the galactic disc, we set a limiting radius of $3.6~\mbox{kpc}$ for integration to exclude orbits within the central area. We chose the value of $3.6~\mbox{kpc}$ because it represents the beginning radius of the spiral structure (Table \ref{parameters}). If the galactocentric distance of a star drops below $3.6~\mbox{kpc}$ during the simulation, the integration immediately stops. We analysed unfinished objects in Sec. \ref{unfinished}. For integration, we used the SciPy \citep{2020SciPy} implementation of the explicit Runge-Kutta method of order 5(4). We measured the time evolution of the radial distance, $R(t)$, with a resolution of $10~\mbox{Myr}$, a guiding radius of $R_g(t)$ with a resolution of $100~\mbox{Myr}$ and the rotation curve $v_{c}(R,t)$ with a resolution of $250~\mbox{Myr}$. To calculate the guiding radius, we adopted a method introduced by \cite{2015A&A...578A..58H}, which consists of linear interpolation between all local maxima and minima for each particle, obtaining the relations $R_{max}(t)$ and $R_{min}(t)$, respectively. Then, the guiding radius was established as the average value:
\begin{equation}\label{R_gyr_Halle}
R_g(t)=\frac{R_{max}(t) + R_{min}(t)}{2}~.
\end{equation}

\subsection{Initial conditions}
We integrated $2 \cdot 10^6$ stars for each simulation with initial positions distributed in $5$ kpc $< R < 20$ kpc. The number of stars at each radius decreases exponentially according to the thin disc density distribution, $\Sigma(R)$, as is described in Sec. \ref{secDisc}. The initial angle, $\varphi$, of each object was chosen randomly.

To determine initial velocities, we used the azimuthal velocity, $v_{\varphi}(R)$, and the radial velocity, $v_{R}(R)$, from \cite{2019A&A...621A..48L}. We added random noise from the interval $(-10,10)~\mbox{km}~\mbox{s}^{-1}$ for $v_{\varphi}$ and $(-5,5)~\mbox{km}~\mbox{s}^{-1}$ for $v_R$. For consistency, we used the same rotation curve for both the DM configuration and the MOND configuration, although the theoretical rotation curves differ a bit (the MOND rotation curve is slightly lower). During the simulation, $v_{\varphi}(R)$ converged to their respective theoretical values in both configurations, the DM case corresponding to the values of \cite{2019A&A...621A..48L}.

A new set of initial conditions was generated for each run (a different set of model parameters, $\Omega_{sp}$ and $A_{sp}$, see Sec. \ref{Num-int}). The DM and MOND configurations were simulated with the same initial conditions.

\subsection{Exclusion of objects in the central region}
\label{unfinished}

We excluded all stars whose radial distance drops below $3.6$ kpc from the simulation, since we are not investigating the central area of the Galaxy. The percentages of such stars are given in Table \ref{tab_unfinished} for each simulation run. Generally, the DM configuration results in far fewer rejected objects than the MOND configuration. In addition, increasing both the spiral amplitude, $A_{sp}$, and the spiral pattern speed, $\Omega_{sp}$, leads to a higher fraction of excluded stars. The initial velocity distributions of the unfinished objects are approximately uniform, unlike the initial positions. The majority of the excluded particles had an initial radius below $6~\mbox{kpc}$ for the DM configuration and below $7~\mbox{kpc}$ for the MOND configuration. The initial angle distributions show a moderate dependency that corresponds to the initial position of the galactic bar. Most of the objects that did not complete the integration passed below the threshold during the first $10^{9}$ years of the simulation.

\begin{table}[!h]
\caption{\centering Unfinished Particles.}
\label{tab_unfinished}
    \centering
    {\small
    \begin{tabular}{|l|c|}\hline
                        simulation run & fraction of unfinished particles [\%] \\ \hline\hline
   DM\_19\_1000 & 7.6 \\
   DM\_25\_850 & 9.0 \\
   DM\_25\_1000 & 11.0 \\
   DM\_25\_1200 & 14.6 \\
   DM\_31\_1000 & 20.3 \\
   \hline
   MOND\_19\_1000 & 11.9 \\
   MOND\_25\_850 & 11.8 \\
   MOND\_25\_1000 & 15.6 \\
   MOND\_25\_1200 & 20.4 \\
   MOND\_31\_1000 & 24.6 \\\hline
                \end{tabular}
    }
\tablefoot{Percentages of particles that did not finish the complete integration for different simulation runs. The particles were removed when they passed below the threshold of $3.6~\mbox{kpc}$.}
\end{table}

\section{Resonances}\label{Resonances}

Using the models described in Sec. \ref{Models}, we calculated the co-rotation and Lindblad resonances for the DM and the MOND model. Based on the small perturbation approximation and the theory of \citet[chap.~3.3]{binney-tremaine2008}, inner Lindblad resonance (ILR) occurs when $\Omega(R)-\Omega_{P} = \frac{\kappa}{m}$, co-rotation resonance occurs when $\Omega(R)-\Omega_{P} = 0$, and OLR occurs when $\Omega(R)-\Omega_{P} = -\frac{\kappa}{m}$, where $m$ is a whole number, either $2$ or $4$. The epicyclic frequency, $\kappa$, is given by
\begin{equation}\label{kappa}
\kappa^{2}=R\frac{\mbox{d}\Omega^{2}}{\mbox{d}R}+4\Omega^{2}~,
\end{equation}
where $\Omega$ is the azimuthal frequency of a star at distance $R$, and $\Omega_{P}$ is the frequency of either the bar or the spiral pattern \citep[][p.~189, eq.~3.139]{binney-tremaine2008}. Higher orders of the co-rotation resonances are denoted as $p:m$ \citep[studied in e.g.][]{lepine2011a,Bautista2021,Trick2021}. They occur when $\Omega(R)-\frac{p}{m}\Omega_{P} = 0$, where $m$ and $p$ are small natural numbers. If the integration period is higher than $m \cdot p \cdot (rotation~period)$, even higher-order resonances might influence radial migration.

The formula for co-rotation resonance using the MOND model follows from Eq. \ref{nonNewt},
\begin{gather}\label{nonNewtOmega}
\begin{aligned}
\frac{v_{c,M}^{2}}{R} = \frac{\vec{g}_{bar}}{1-\mbox{exp}(-\sqrt{g_{bar}/g_{\dag}})} \equiv \frac{\Omega_{0,M}^{2} R^2}{R}~,
\end{aligned}
\end{gather}
where $v_{c,M}$ is the circular velocity at distance $R$, $\Omega_{0,M}$ is the azimuthal frequency at distance $R$, and $g_{bar}$ is defined by Eq. \ref{nonNewt}.
The epicyclic frequency can then be calculated from Eq. \ref{kappa} as
\begin{equation}\label{nonNewtkappa}
\kappa_{M}^{2} = R \frac{\mbox{d} \Omega_{0,M}^{2}}{\mbox{d} R} + 4 \Omega_{0,M}^{2}~.
\end{equation}

\begin{figure*}
\centering 

\subfloat{%
  \includegraphics[width=1\textwidth]{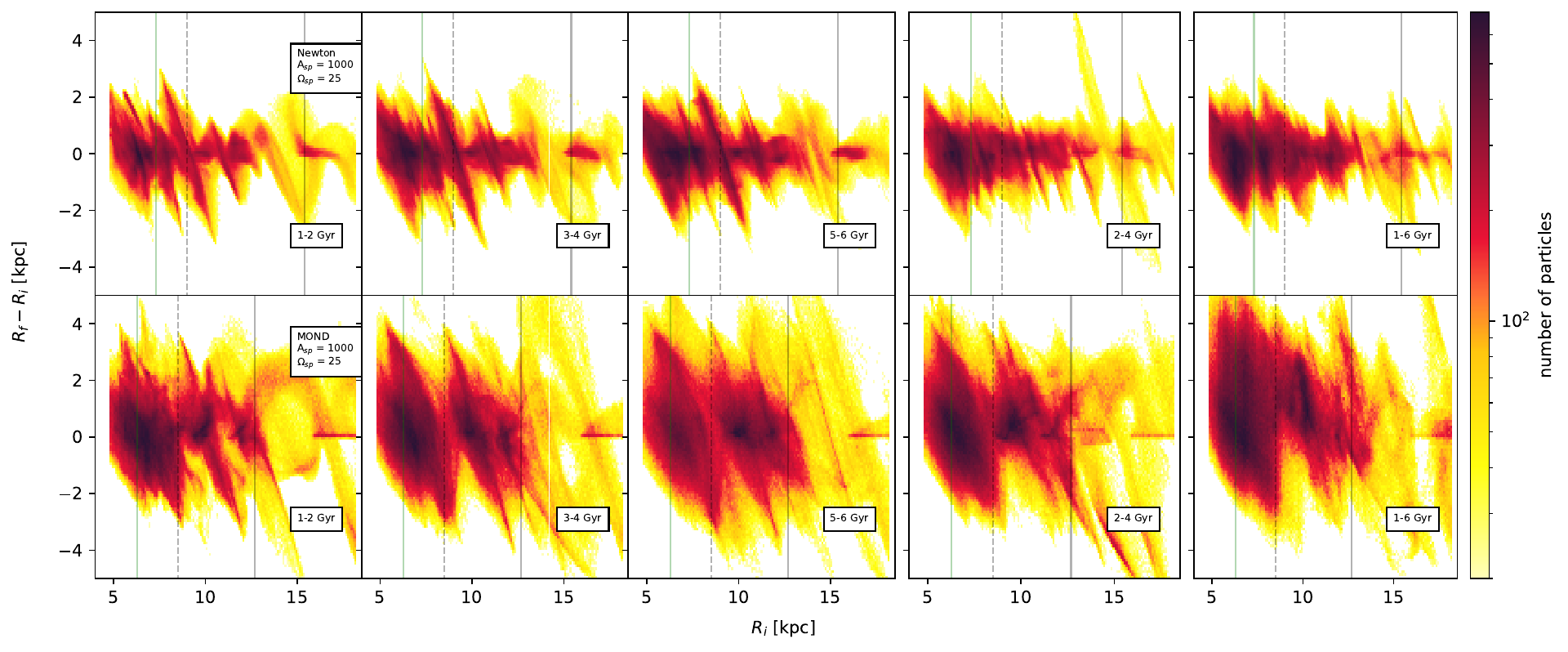}%
}

\caption{Distribution of changes in galactocentric radius, $R_{f} - R_{i}$, as a function of $R_{i}$ for different time intervals during the $6~\mbox{Gyr}$ integration. The upper panel represents the simulation run DM\_25\_1000; the lower panel represents the simulation run MOND\_25\_1000 (corresponding parameters as listed in Tables \ref{parameters} and \ref{parameters2}). Vertical lines represent theoretical values of resonances: green lines represent resonances produced by the galactic bar, and grey lines represent resonances caused by the spiral arms. In both cases, the solid lines represent OLRs and the dashed line the $m=p=1$ co-rotation resonance.}
\label{Fig22.0}
\end{figure*}

\begin{figure*}
\centering 

\subfloat{%
  \includegraphics[width=.95\textwidth]{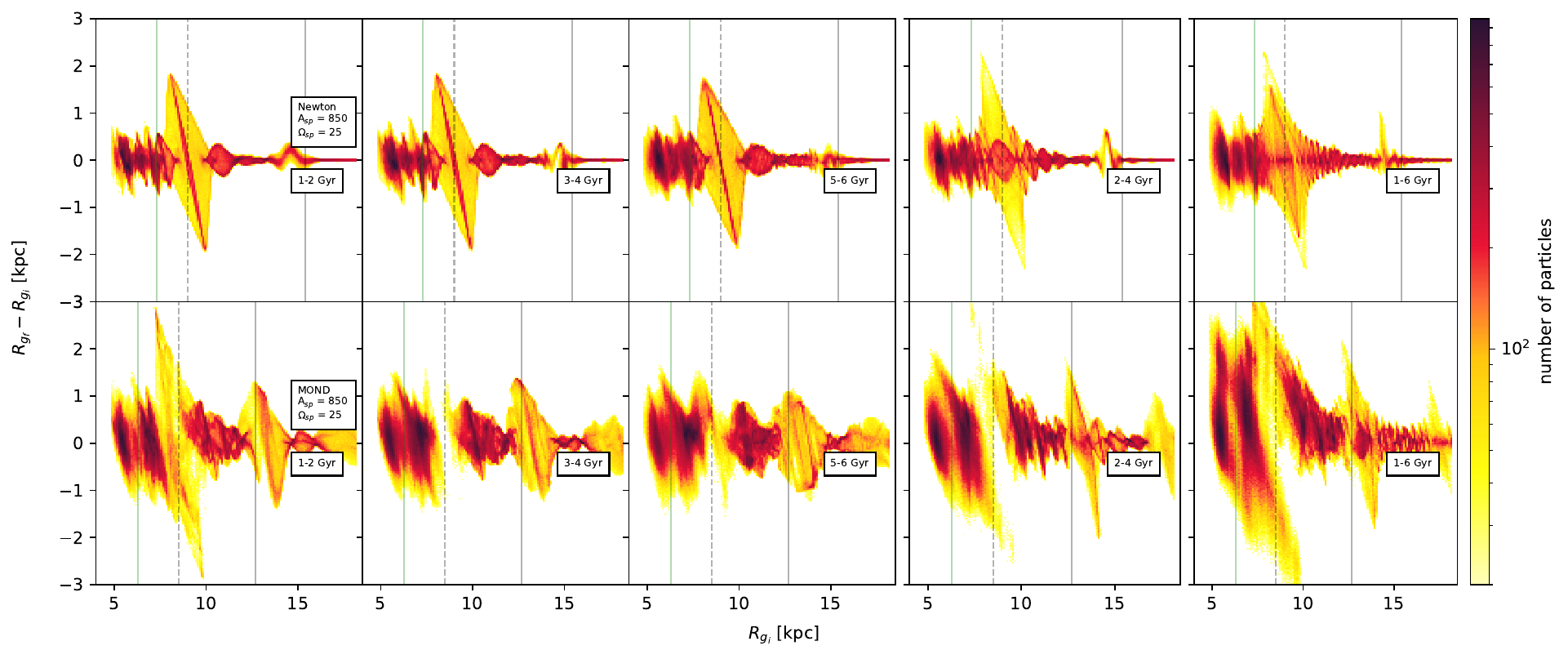}%
}

\subfloat{%
  \includegraphics[width=.95\textwidth]{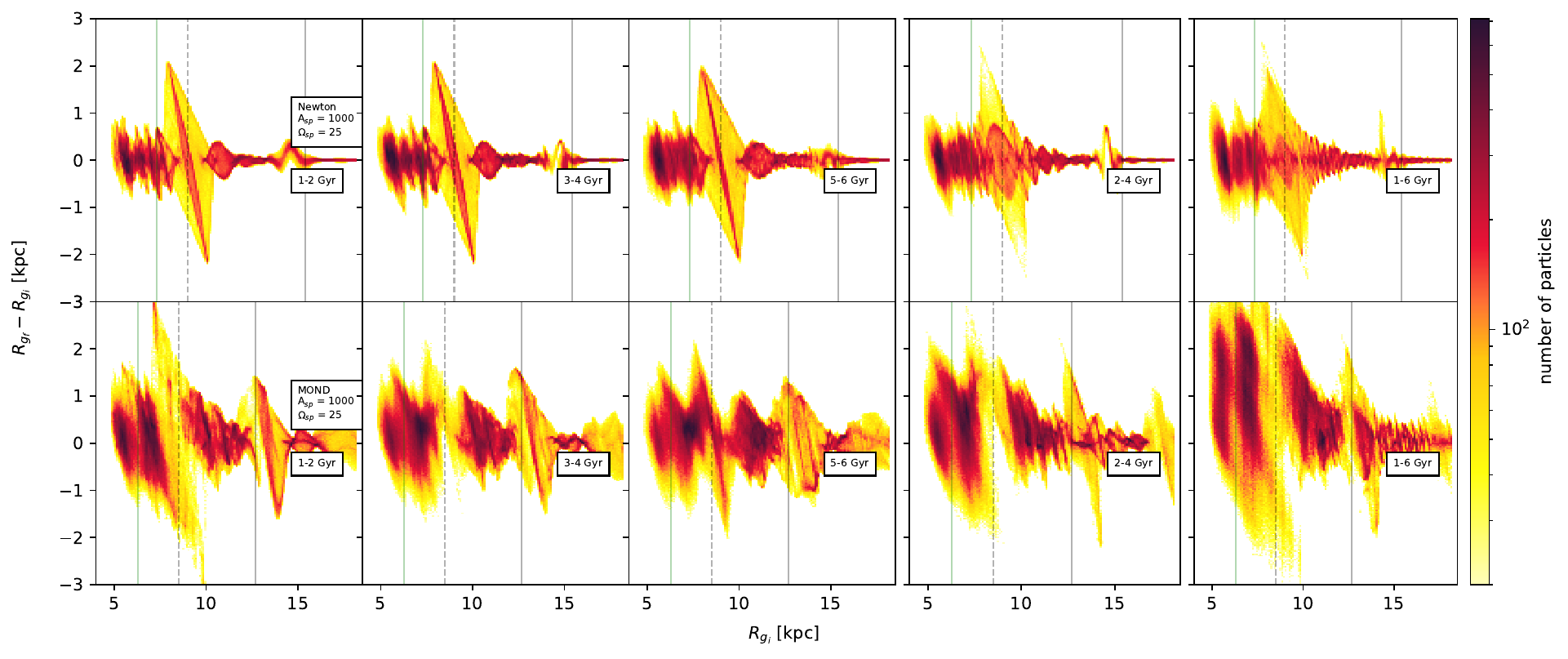}%
}

\subfloat{%
  \includegraphics[width=.95\textwidth]{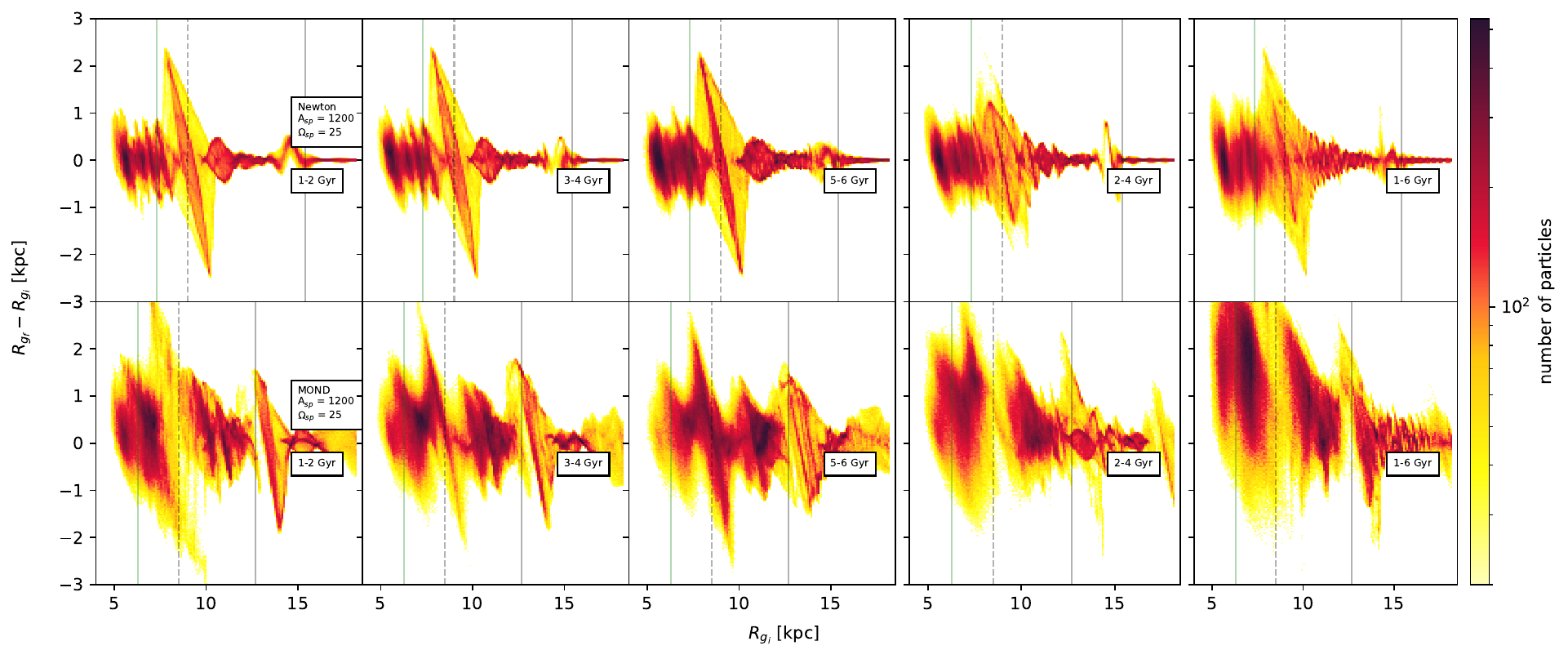}%
}

\caption{Distribution of changes in the guiding radius, $R_{g_{f}} - R_{g_{i}}$, 
(radial migration) as a function of $R_{g_{i}}$ for different time intervals during the $6~\mbox{Gyr}$ integration. Each set of figures represents a different simulation run with parameters as listed in Tables \ref{parameters} and \ref{parameters2}. Vertical lines represent the theoretical values of resonances: green lines for resonances produced by the galactic bar and grey lines for resonances caused by the spiral arms. In both cases, the solid lines represent OLRs and the dashed line the $m=p=1$ co-rotation resonance.}
\label{Fig11}
\end{figure*}

\begin{figure*}
\centering

\subfloat{%
  \includegraphics[width=.95\textwidth]{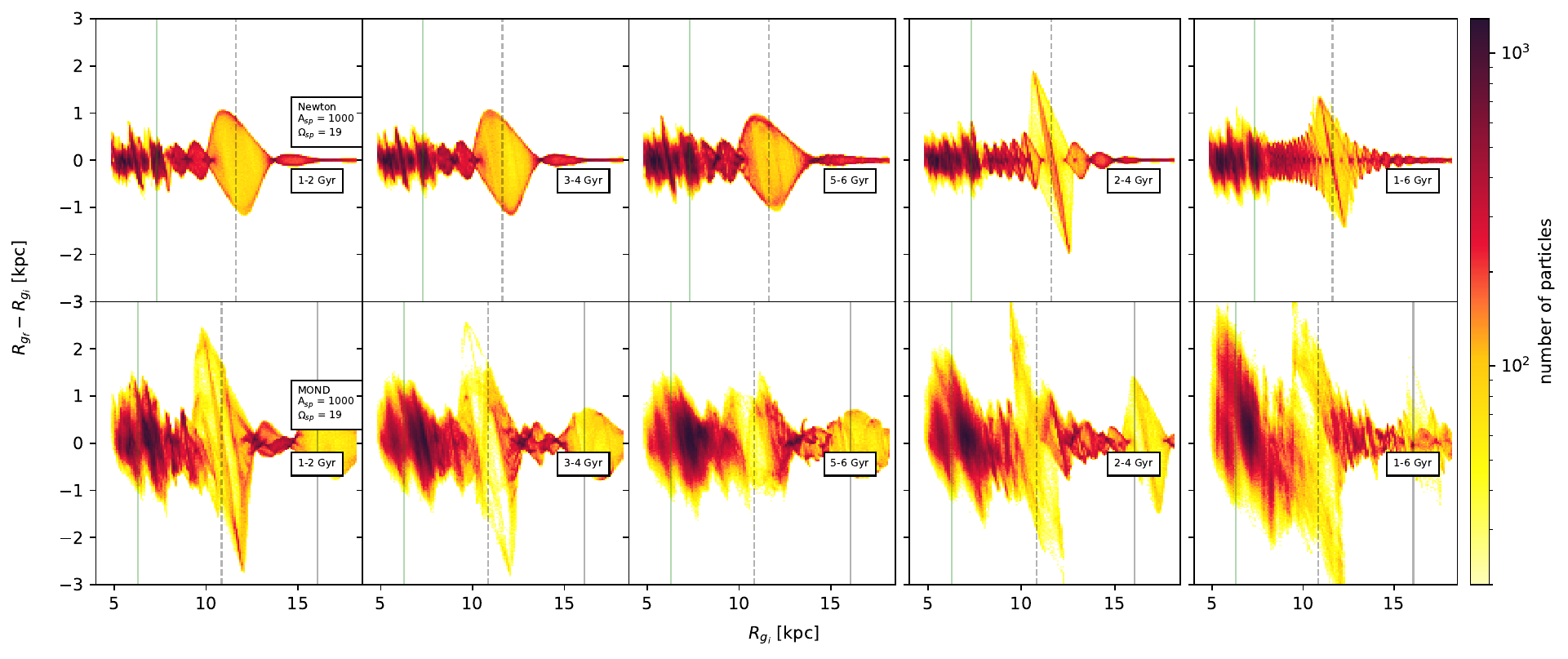}%
}

\subfloat{%
  \includegraphics[width=.95\textwidth]{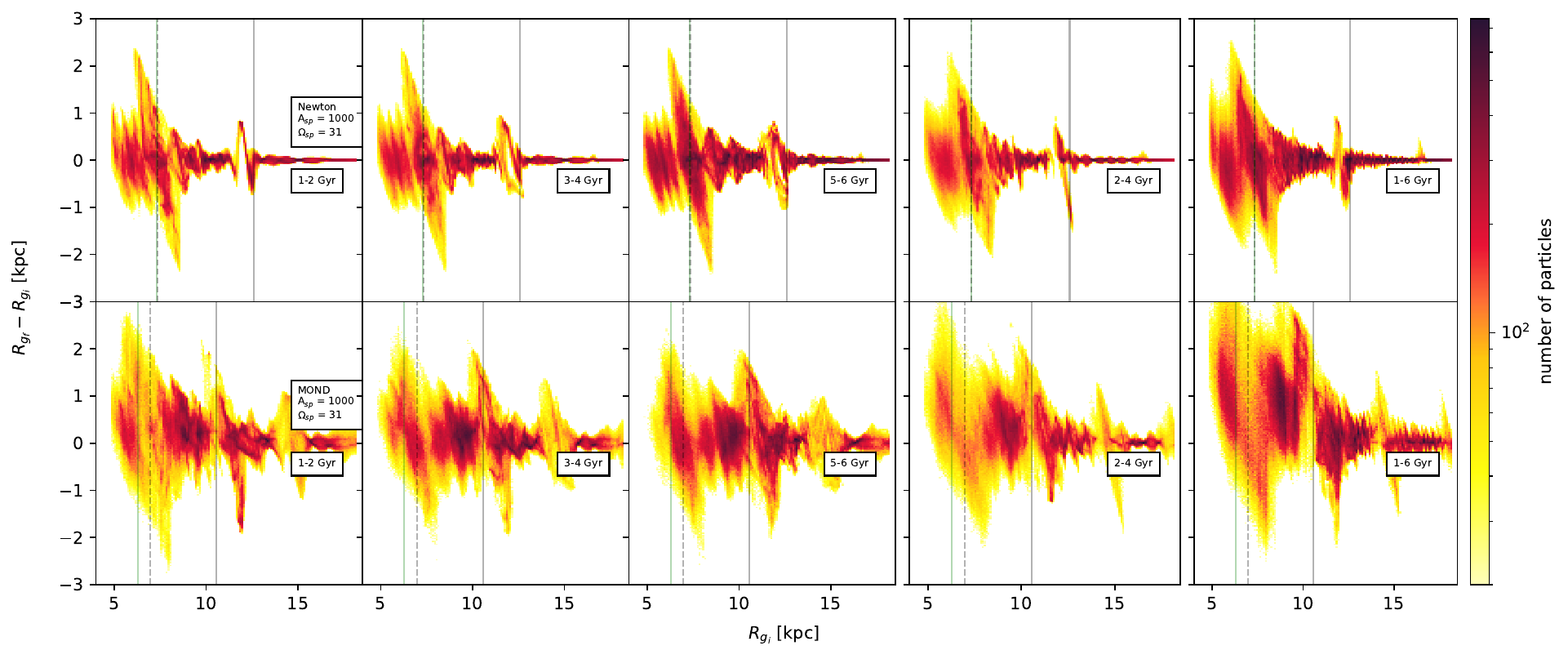}%
}
\caption{Continuation of Fig. \ref{Fig11}}
\label{Fig11.2}
\end{figure*}

\subsection{Resonances in the frequency phase space}

To analyse resonances of our models in more detail, we investigated the frequency phase space \citep[e.g.][]{Ceverino2007}, which enabled us to directly determine which resonances are present in each model as well as their relative strength and time evolution. This method is more appropriate than the epicyclic approximation that assumes nearly circular orbits, which is not always the case in our simulations (especially for lower radial distances). 

If we consider a particle rotating around the centre with an azimuthal frequency, $\Omega_{\varphi} = \Omega$, and radial frequency, $\Omega_R = \kappa$, then the resonance occurs when
\begin{equation}\label{eq_resonance}
l\cdot \kappa + m\cdot \Omega = p\cdot \Omega_P~,
\end{equation}
where $\Omega_P$ is the pattern speed of the bar or the spiral arms, and $l$, $m$, and $p$ are small integers. In this approach, co-rotation resonances occur when $l=0$ (sometimes denoted as $p:m$), while the ILR corresponds to $l=-1$, $m=p=2$ and the OLR to $l=1$, $m=p=2$. The frequency phase space consists of the azimuthal frequency on the horizontal axis and the radial frequency on the vertical axis. Thus, co-rotation resonances appear as vertical lines in the frequency phase space. Each additional resonance will appear as a line in the frequency space. If a star is trapped in a resonance, its orbit is closed in a reference frame rotating with the pattern. Such periodic orbits may lead to a concentration of stars around resonance lines in frequency space.

Assuming linearisation of the rotation curve \citep[e.g.][]{Eilers2019} in the form $v_c(R) = A\cdot R + B$, we also calculated an approximate epicyclic curve in the frequency phase space as
\begin{equation}\label{eq_epiapprox}
\kappa(\Omega) = \sqrt{2} \cdot\Omega\sqrt{1+\frac{A}{\Omega}}~.
\end{equation}

\subsection{Comparison of resonance positions}

\begin{table*}
\caption{\centering Resonance positions.}
\label{TableMAXI}
\centering
\def\arraystretch{1.3}
\begin{tabular}{ |c|c||c|c||c|c||c|c||c|c| } 
 \hline
 \multicolumn{2}{|c||}{Structure} & \multicolumn{2}{c||}{Bar} & \multicolumn{6}{c|}{Spiral arms}\\
 \cline{1-10}
  \multicolumn{2}{|c||}{$\Omega~[\mbox{km}~\mbox{s}^{-1}~\mbox{kpc}^{-1}]$} & \multicolumn{2}{c||}{55.5} & \multicolumn{2}{c||}{$19$} & \multicolumn{2}{c||}{$25$} & \multicolumn{2}{c|}{$31$}\\
 \cline{1-10}
  \multicolumn{2}{|c||}{Regime} & DM & MOND & DM & MOND & DM & MOND& DM & MOND\\
 \hline
 \hline
 \multirow{2}{*}[-0.2em]{Lindblad $2:1$} & Inner & -- & -- & -- & 4.56 & -- & 3.68 & -- & 3.17\\
  & Outer & 7.31 & 6.28 & 19.99 & 16.07 & 15.40 & 12.68 & 12.59 & 10.56\\
 \hline
 \multirow{2}{*}[-0.2em]{Lindblad $3:1$} & Inner & -- & -- & 5.72 & 6.78 & 4.33 & 5.18 & 3.49 & 4.23\\
  & Outer & 6.27 & 5.51 & 17.22 & 14.35 & 13.28 & 11.35 & 10.88 & 9.43\\
 \hline
 \multirow{2}{*}[-0.2em]{Lindblad $4:1$} & Inner & -- & -- & 7.28 & 7.89 & 5.55 & 6.04 & 4.45 & 4.89\\
  & Outer & 5.74 & 5.11 & 15.62 & 13.49 & 12.22 & 10.67 & 10.01 & 8.85\\
 \hline
 \multirow{3}{*}[-0.08em]{Co-rotation} & $1:1$ & 4.09 & 3.90 & 11.62 & 10.84 & 9.00 & 8.50 & 7.34 & 6.96\\
  & $5:3$ & 6.85 & 6.50 & 18.81 & 16.83 & 14.49 & 13.29 & 11.85 & 11.04\\
  & $7:5$ & 5.77 & 5.49 & 15.92 & 14.48 & 12.31 & 11.44 & 10.08 & 9.48\\
 \hline
\end{tabular}
\tablefoot{Calculated galactocentric distances of Lindblad and co-rotation resonances, in kiloparsecs. In the case of the Lindblad resonances, the value represents the centre of the resonance zone. Empty cells are outside of the range of $(3-20)~\mbox{kpc}$.}
\end{table*}
  
\begin{table*}\label{Resonances_previous_results}
\caption{\centering Comparison of resonances positions.}
\label{tab_comparison}
    \centering
    {\small
    \begin{tabular}{|c||c|c|c|} \hline
         Work & Resonance type & Pattern speed [$\mbox{km}~\mbox{s}^{-1}~\mbox{kpc}^{-1}$] & Position [$\mbox{kpc}$] \\ \hline \hline
         \citet{quillen-minchev2005} & ILR4:1 spiral & 18.1 & 8.0 \\ \hline
         \citet{antoja2011} & ILR4:1 spiral & 18 & 8.0 \\ \hline
         \citet{lepine2011a} & CR1:1 spiral & 24 & 8.4 \\
         & OLR2:1 spiral & 24 & 14.0 \\
         & ILR4:1 spiral & 24 & 6.2 \\ \hline
         \citet{martinez-barbosa2015} & CR1:1 spiral & 20 & 10.9 \\
          & OLR2:1 spiral & 20 & 16.0 \\
          & OLR2:1 bar & 40 & 10.2 \\ \hline
         \citet{Bovy2019} & CR1:1 bar & 41 & 5.5 \\
         & OLR4:1 bar & 41 & 7.7 \\
         & OLR2:1 bar & 41 & 9.5 \\ \hline
         \cite{Lucchini2024} & OLR2:1 bar & 40 & 12 \\ \hline
    \end{tabular}   }
    \tablefoot{Comparison of the position of various resonances in previously published works.}
\end{table*}

In Table \ref{TableMAXI}, we summarise the regions of co-rotation and Lindblad resonances. We consider different values of the spiral pattern speed -- low, middle, and high values.

In Table \ref{tab_comparison}, we present the positions of the resonances published in the literature. Our results are in good agreement with previous works that use similar values of the spiral pattern speed. The differences between positions of resonances induced by the bar are due to different pattern speeds considered in other works. 

\section{Results}\label{Results}

\subsection{Radial migration}
\label{51radmig}

To analyse results from our simulations, we investigated the radial migration (the change of guiding radius) and the overall change in the galactocentric radius as well. In Figs. \ref{Fig22.0}, \ref{Fig22}, and \ref{Fig22.2}, we show the change of the final and initial galactocentric radius, $R_f - R_i$, as a function of the initial radius, $R_i$, for different time intervals during the $6~\mbox{Gyr}$ integration for various settings of $A_{sp}$ and $\Omega_{sp}$ in both the DM and MOND configurations (details about used parameters for all ten simulation runs are given in Tables \ref{parameters} and \ref{parameters2}). Similarly, Figs. \ref{Fig11} and \ref{Fig11.2} show the change in the final and initial guiding radius, $R_{g_{f}} - R_{g_{i}}$, as a function of the initial guiding radius, $R_{g_{i}}$.

The change in the galactocentric radius is observable throughout the simulation (see Figs. \ref{Fig22.0}, \ref{Fig22} and \ref{Fig22.2}), although this effect is more prominent during the first half of the simulation, a result consistent with \citet{halle2018}. Early on, the disc of the galaxy is less stabilised, which allows for more significant interactions and perturbations between stars and the galactic bar and spiral arms. As the galaxy evolves, these effects contribute to an overall increase in the dispersion of stellar orbits, demonstrating how the change in the galactocentric radius shapes the structure of the galactic disc and the stellar distribution.

As was expected, in the comparative analysis of the DM model versus the MOND one, the effect of resonances and the changes in both the galactocentric radius and guiding radius are significantly more pronounced under the MOND framework. This is because when the MOND model is used, the non-axisymmetric galactic components responsible for the resonances represent a larger portion of the total gravitational action of the MW.

The influence of the spiral arms parameters (amplitude, $A_{sp}$, and pattern speed, $\Omega_{sp}$) on the system is apparent, shaping the dynamics of the disc. As the amplitude of the spiral pattern, $A_{sp}$, increases, it enhances the gravitational torque exerted on the stars, leading to a higher relative strength of resonances and more pronounced effects in both the radial migration and the change in the galactocentric radius. As was expected, higher amplitudes lead to more extensive shifts in stellar orbits and increase the spread of orbital eccentricities across the galactic disc.

The co-rotation resonance, $m=p=1$, with the spiral arms is especially important in enhancing the migration effects. Its impact on stellar orbits is clearly observable. Together with other resonances (especially the OLR with the bar and spiral arms), their collective influence initiates a complex migration that significantly impacts the structural evolution of the galaxy and the spatial distribution of its stellar components. The difference between positions of the resonances is mainly due to the difference between the rotation curves for Newtonian gravity and MOND. It can be seen from Eq. \ref{nonNewtOmega} and Eq. \ref{nonNewtkappa} that the position of the resonances is effectively a function of the rotation curve. Therefore, the differences in the rotation curves explain the differences in the position of the resonance-induced peaks in Figs. \ref{Fig22.0}, \ref{Fig22}, \ref{Fig22.2}, \ref{Fig11}, and \ref{Fig11.2}.

The feature visible in the first three panels of the top row of Fig. \ref{Fig11.2} at the spiral arms' co-rotation resonance is caused by the short time interval for the low value of the spiral pattern speed, $\Omega_{sp} = 19~\mbox{km}~\mbox{s}^{-1}~\mbox{kpc}^{-1}$. The $1~\mbox{Gyr}$ time interval represents only about three revolutions of the spiral arms; thus, the effect of the resonance does not have enough time to fully manifest. For longer time intervals, this feature vanishes.

\subsection{Resonance analysis} \label{reson}

We investigated the frequency phase space and its time evolution for all three $\Omega_{sp}$ configurations, tracking the average orbital and radial frequency of each particle with a $1~\mbox{Gyr}$ time step. In general, the most prominent resonances in all five configurations are the spiral arms' co-rotation resonance, $m=p=1$, the bar and the spiral arms' OLR, and the co-rotation resonance, $m=2, p=1$, with the superposition of the galactic bar and spiral arms, $2 \Omega = \Omega_b + \Omega_{sp}$. The galactic bar co-rotation resonance, $m=p=1$, which corresponds to pattern speed of $55.5~\mbox{km}~\mbox{s}^{-1}~\mbox{kpc}^{-1}$, is not present, as it lies beyond the range of our phase space. The epicycle approximation curve reasonably captures the general trend for lower frequencies but underestimates it for higher frequencies. This is not surprising, as these areas correspond to a lower radial distance, where the actual rotation curve deviates more from the linear approximation of the theoretical rotation curve of our model. In addition, areas of smaller radius exhibit orbits that are generally less circular than orbits in more distant parts of the disc. We also found that the frequency space evolves more rapidly over time in the MOND regime than in the DM regime. Remarkably, the prominent spiral arms co-rotation resonance, $m=p=1$, has a tendency to first remove and then pull back stars from its position in all MOND configurations. In contrast, DM configurations produce a stable cluster of objects near the $m=p=1$ spiral arm co-rotation resonance. Another difference between the time evolution of the frequency phase space in the DM and MOND configuration is that the MOND model leads to fluctuations around the epicycle approximation curve of particles with low azimuthal frequencies, while the DM regime displays more steady distributions at low frequencies. Finally, the MOND distributions are systematically shifted to the left (smaller $\Omega$) in the frequency phase space compared to the DM distributions. This is in agreement with both theoretical and simulated rotation curves, which are slightly higher for the DM configuration than for the MOND configuration, resulting in systematically lower orbital frequencies $\Omega$ for MOND. In addition to these general features, we took a closer look at each $\Omega_{sp}$ configuration individually:

\begin{figure*}
\centering
\subfloat{\includegraphics[width=0.98\textwidth]{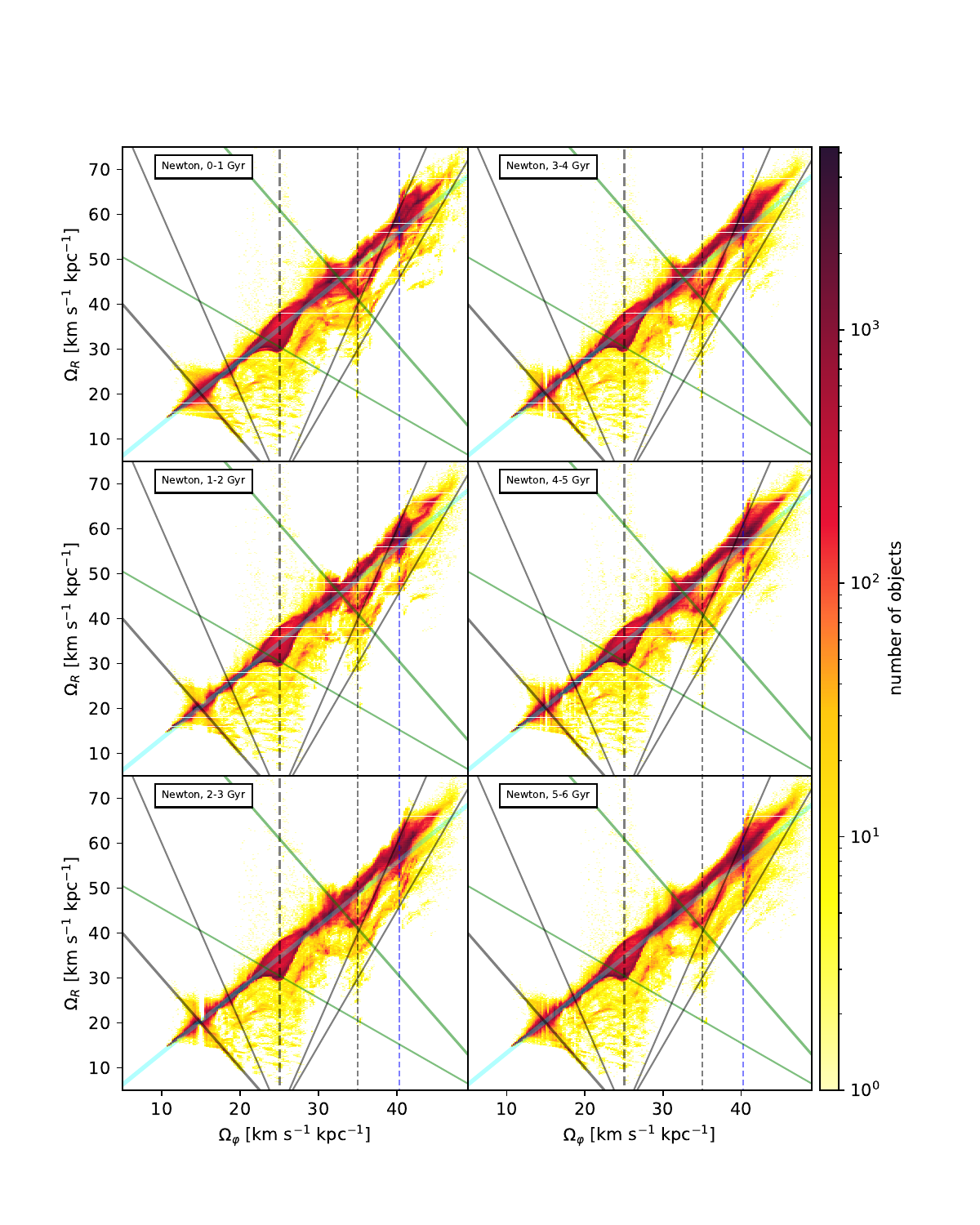}}
\caption{Time evolution of the frequency phase space with a time step of $1~\mbox{Gyr}$ for the simulation run DM\_25\_1000. The light blue curve indicates the epicycle approximation. Displayed resonances include spiral arm co-rotation resonances, $m=p=1$ and $m=5$, $p=7$ (thick and thin dashed grey lines, respectively), the co-rotation resonance $m=2$, $p=1$ with the superposition of the spiral arms and the galactic bar (dashed blue line), the OLR with the spiral arms (thick solid grey line), the OLR with the galactic bar (thick solid green line), a bar resonance of $l=m=p=1$ (thin green line), and spiral arm resonances of $l=-1$, $m=p=4$, $l=-1$, $m=p=3$, and $l=1$, $m=p=4$ (thin solid grey lines).}
\label{freq_02N}
\end{figure*}

\begin{figure*}
\centering
\subfloat{\includegraphics[width=1\textwidth]{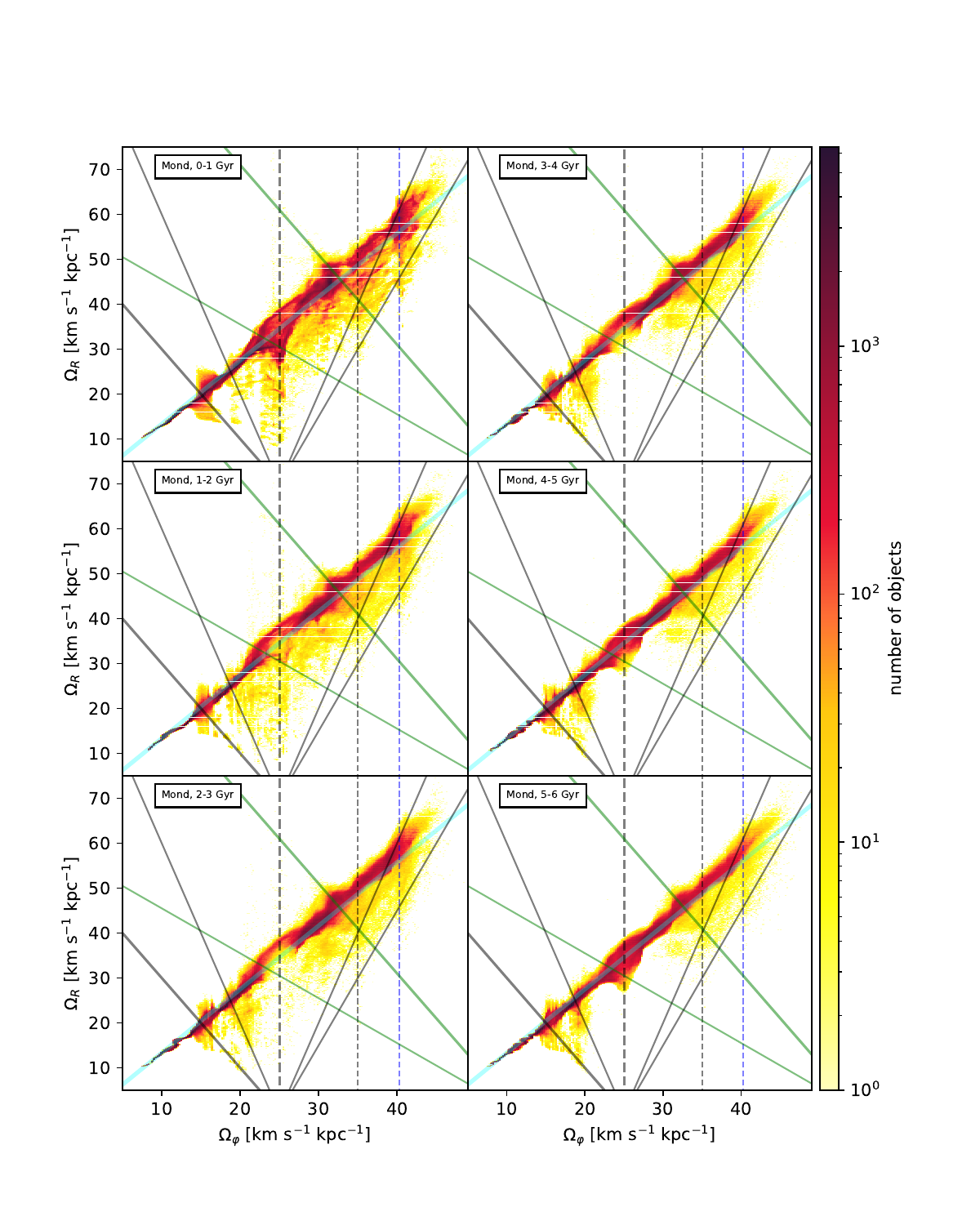}}
\caption{Same as Fig. \ref{freq_02N}, but for the simulation run MOND\_25\_1000.}
\label{freq_02M}
\end{figure*}

\begin{itemize}
    \item[$\bullet$] $\Omega_{sp} = 19 \, \mathrm{km} \, \mathrm{s}^{\mathrm{-1}} \, \mathrm{kpc}^{\mathrm{-1}}$: The frequency maps for simulation runs DM\_19\_1000 and MOND\_19\_1000 are shown in Figs. \ref{freq_01N} and \ref{freq_01M}, respectively. The effect of the OLR from the spiral arms is weakened as the small value of $\Omega_{sp}$ places it towards lower frequencies (larger radial distances). However, this moves the ILR with the spiral arms closer to the epicycle approximation curve. In fact, this is the only configuration in which the ILR with the spiral arms is well populated. Remarkably, it stabilises objects in the frequency phase space. Another prominent line covering a wide interval of frequencies is the spiral arms resonance $l=-1$, $m=p=3$. The spiral arms resonance $l=-1$, $m=p=4$ is also present, most notably in the area where it intersects the spiral arms co-rotation resonance $m=3$, $p=5$ and the OLR with the galactic bar, resulting in a big cluster of particles. When compared to the DM configuration, the MOND configuration exhibits a more prominent co-rotation resonance, $m=2$, $p=1$, with the superposition of the galactic bar and the spiral arms.
    \item[$\bullet$] $\Omega_{sp} = 25 \, \mathrm{km} \, \mathrm{s}^{\mathrm{-1}} \, \mathrm{kpc}^{\mathrm{-1}}$: We simulated configurations with three different $A_{sp}$ for this value of $\Omega_{sp}$ (see Table \ref{parameters2}), but there was only a small quantitative difference. The time evolution of the frequency phase space for simulation runs DM\_25\_1000 and MOND\_25\_1000 are shown in Figs. \ref{freq_02N} and \ref{freq_02M}, respectively. The prominent spiral arms co-rotation resonance $m=p=1$ overlaps with the galactic bar resonance $l=m=p=1$. The co-rotation resonance $m=2$, $p=1$ from the superposition of the spiral arms and the galactic bar, the spiral arms resonances $l=-1$, $m=p=3$, $l=-1$, $m=p=4$, and $l=1$, $m=p=4$, and the spiral arm co-rotation resonance $m=5$, $p=7$ are also visible, especially in the DM configuration.
    \item[$\bullet$] $\Omega_{sp} = 31 \, \mathrm{km} \, \mathrm{s}^{\mathrm{-1}} \, \mathrm{kpc}^{\mathrm{-1}}$: The frequency maps for simulation runs DM\_31\_1000 and MOND\_31\_1000 are shown in Figs. \ref{freq_03N} and \ref{freq_03M}, respectively. The spiral arm co-rotation resonance, $m=p=1$, is very strong in this case. It intersects with the OLR with the galactic bar. The spiral arm resonance, $l=1$, $m=p=4$, is also visible, crossing the galactic bar resonance, $l=m=p=1$. The co-rotation resonance, $m=2$, $p=1$, with the superposition of the galactic bar and spiral arms is present only in the DM configuration.
\end{itemize}

\subsection{Maximal change in the guiding radius -- A statistical analysis}
\label{extrmni migranti}

\begin{table*}

\caption{\centering The maximal change of guiding radii in our simulations for various parameters of the spiral structure.}
\label{Tab_radmig}
\resizebox{\textwidth}{!}{
\centering
\def\arraystretch{1.3}

\begin{tabular}{|c|c||c||c|c||c|c||c|c||c|c||c|c||c|c||c|c|}

 \hline
 $\Omega_{sp}$ & $A_{sp}$ & $\mbox{d}R_{g}~[\mbox{kpc}]$ & \multicolumn{2}{c||}{$0-1~\mbox{Gyr}$} & \multicolumn{2}{c||}{$1-2~\mbox{Gyr}$} & \multicolumn{2}{c||}{$2-3~\mbox{Gyr}$} & \multicolumn{2}{c||}{$3-4~\mbox{Gyr}$} & \multicolumn{2}{c||}{$4-5~\mbox{Gyr}$} & \multicolumn{2}{c||}{$5-6~\mbox{Gyr}$} & \multicolumn{2}{c|}{$2-6~\mbox{Gyr}$}\\
 \hline
 &&& DM & MOND & DM & MOND & DM & MOND & DM & MOND & DM & MOND & DM & MOND & DM & MOND\\
 \hline
 \hline
 19 & 1000 & <-2.5 & 0.00 & 0.00 & 0.00 & 0.02 & 0.00 & 0.03 & 0.00 & 0.05 & 0.00 & 0.02 & 0.00 & 0.00 & 0.00 & 2.13\\
 &  & (-2.5, -1.5) & 0.00 & 0.74 & 0.00 & 2.20 & 0.00 & 1.11 & 0.00 & 0.90 & 0.00 & 0.29 & 0.00 & 0.10 & 2.20 & 5.56\\
 &  & (-1.5, -0.5) & 7.51 & 33.94 & 7.37 & 32.74 & 7.40 & 30.57 & 8.16 & 30.10 & 7.85 & 28.12 & 8.24 & 27.00 & 22.41 & 27.72\\
 &  & (-0.5, 0.5) & 83.74 & 29.83 & 85.98 & 30.11 & 85.57 & 32.80 & 84.95 & 34.58 & 84.58 & 37.88 & 84.50 & 41.02 & 51.78 & 11.36\\
 &  & (0.5, 1.5) & 8.75 & 33.91 & 6.65 & 33.66 & 7.03 & 34.01 & 6.88 & 33.88 & 7.57 & 33.45 & 7.26 & 31.80 & 19.88 & 39.14\\
 &  & (1.5, 2.5) & 0.00 & 0.00 & 0.00 & 1.27 & 0.00 & 1.48 & 0.00 & 0.50 & 0.00 & 0.24 & 0.00 & 0.07 & 1.80 & 11.85\\
 &  & >2.5 & 0.00 & 0.00 & 0.00 & 0.00 & 0.00 & 0.00 & 0.00 & 0.00 & 0.00 & 0.00 & 0.00 & 0.00 & 0.00 & 2.24\\
 \hline
 25 & 850 & <-2.5 & 0.00 & 0.05 & 0.00 & 0.25 & 0.00 & 0.03 & 0.00 & 0.00 & 0.00 & 0.00 & 0.00 & 0.00 & 0.00 & 0.66\\
 &  & (-2.5, -1.5) & 0.23 & 3.37 & 0.80 & 2.30 & 0.69 & 0.59 & 0.72 & 0.09 & 0.74 & 0.02 & 0.65 & 0.05 & 4.15 & 6.17\\
 &  & (-1.5, -0.5) & 17.48 & 29.59 & 18.02 & 29.76 & 16.45 & 30.66 & 19.11 & 26.27 & 17.57 & 23.34 & 18.86 & 21.46 & 30.73 & 18.40\\
 &  & (-0.5, 0.5) & 62.64 & 31.34 & 65.21 & 31.52 & 63.98 & 34.48 & 63.59 & 37.60 & 63.66 & 41.17 & 63.03 & 44.39 & 32.51 & 12.49\\
 &  & (0.5, 1.5) & 19.49 & 31.34 & 15.57 & 33.66 & 18.36 & 33.77 & 16.11 & 35.96 & 17.58 & 35.40 & 17.03 & 34.02 & 28.16 & 47.07\\
 &  & (1.5, 2.5) & 0.16 & 3.98 & 0.40 & 2.24 & 0.53 & 0.42 & 0.47 & 0.08 & 0.45 & 0.07 & 0.43 & 0.08 & 4.45 & 14.62\\
 &  & >2.5 & 0.00 & 0.34 & 0.00 & 0.26 & 0.00 & 0.04 & 0.00 & 0.00 & 0.00 & 0.00 & 0.00 & 0.00 & 0.00 & 0.58\\
 \hline
 25 & 1000 & <-2.5 & 0.00 & 0.33 & 0.00 & 0.37 & 0.00 & 0.03 & 0.00 & 0.00 & 0.00 & 0.00 & 0.00 & 0.00 & 0.07 & 0.63\\
 &  & (-2.5, -1.5) & 0.86 & 5.02 & 1.74 & 2.40 & 1.45 & 0.75 & 1.68 & 0.24 & 1.56 & 0.24 & 1.54 & 0.46 & 5.01 & 6.35\\
 &  & (-1.5, -0.5) & 22.04 & 30.53 & 23.78 & 30.61 & 22.52 & 29.53 & 24.02 & 25.43 & 22.90 & 24.56 & 24.96 & 23.92 & 32.49 & 13.81\\
 &  & (-0.5, 0.5) & 50.68 & 24.49 & 52.47 & 23.87 & 51.22 & 26.86 & 51.00 & 30.95 & 51.21 & 33.60 & 50.12 & 36.59 & 26.97 & 10.70\\
 &  & (0.5, 1.5) & 25.45 & 34.09 & 20.90 & 39.72 & 23.35 & 41.97 & 22.09 & 42.77 & 22.94 & 41.03 & 22.17 & 38.36 & 30.20 & 34.80\\
 &  & (1.5, 2.5) & 0.97 & 4.69 & 1.11 & 2.43 & 1.46 & 0.81 & 1.21 & 0.59 & 1.39 & 0.57 & 1.20 & 0.68 & 5.17 & 30.36\\
 &  & >2.5 & 0.00 & 0.85 & 0.00 & 0.60 & 0.00 & 0.06 & 0.00 & 0.01 & 0.00 & 0.00 & 0.00 & 0.00 & 0.08 & 3.34\\
 \hline
 25 & 1200 & <-2.5 & 0.00 & 1.42 & 0.00 & 0.59 & 0.00 & 0.10 & 0.00 & 0.05 & 0.00 & 0.05 & 0.00 & 0.24 & 1.79 & 1.48\\
 &  & (-2.5, -1.5) & 1.88 & 7.06 & 3.22 & 3.26 & 2.83 & 1.38 & 2.90 & 1.18 & 3.04 & 1.67 & 2.78 & 2.26 & 4.91 & 6.34\\
 &  & (-1.5, -0.5) & 27.02 & 28.40 & 28.25 & 28.40 & 27.73 & 25.95 & 28.70 & 25.76 & 28.00 & 27.38 & 29.86 & 27.83 & 34.29 & 12.19\\
 &  & (-0.5, 0.5) & 39.92 & 18.75 & 40.57 & 18.73 & 39.78 & 21.78 & 39.51 & 23.54 & 39.31 & 25.27 & 38.66 & 26.75 & 22.35 & 8.85\\
 &  & (0.5, 1.5) & 28.96 & 36.86 & 25.68 & 44.05 & 26.86 & 47.58 & 26.24 & 46.68 & 27.15 & 42.92 & 26.17 & 39.52 & 30.29 & 25.27\\
 &  & (1.5, 2.5) & 2.22 & 6.21 & 2.28 & 3.99 & 2.79 & 2.92 & 2.65 & 2.63 & 2.50 & 2.59 & 2.53 & 3.23 & 4.58 & 32.72\\
 &  & >2.5 & 0.00 & 1.29 & 0.00 & 0.98 & 0.00 & 0.29 & 0.00 & 0.16 & 0.00 & 0.13 & 0.00 & 0.18 & 1.8 & 13.14\\
 \hline
 31 & 1000 & <-2.5 & 0.00 & 0.56 & 0.00 & 0.34 & 0.00 & 0.13 & 0.00 & 0.08 & 0.00 & 0.11 & 0.00 & 0.18 & 0.23 & 1.80\\
 &  & (-2.5, -1.5) & 3.81 & 6.33 & 1.97 & 3.08 & 2.34 & 1.63 & 2.85 & 1.95 & 3.00 & 2.35 & 2.64 & 2.58 & 10.11 & 7.31\\
 &  & (-1.5, -0.5) & 22.46 & 16.65 & 26.03 & 22.48 & 25.32 & 24.72 & 23.66 & 25.95 & 23.59 & 26.11 & 24.00 & 26.66 & 21.32 & 13.96\\
 &  & (-0.5, 0.5) & 42.30 & 29.01 & 43.60 & 28.79 & 45.20 & 32.63 & 46.18 & 33.86 & 46.35 & 34.81 & 46.93 & 35.24 & 35.16 & 16.55\\
 &  & (0.5, 1.5) & 27.91 & 36.74 & 23.72 & 39.93 & 24.55 & 37.27 & 24.55 & 34.93 & 24.41 & 33.10 & 23.68 & 32.00 & 23.02 & 30.89\\
 &  & (1.5, 2.5) & 3.52 & 9.99 & 4.68 & 4.57 & 2.59 & 3.26 & 2.76 & 3.04 & 2.66 & 3.37 & 2.75 & 3.18 & 9.82 & 22.9\\
 &  & >2.5 & 0.00 & 0.72 & 0.00 & 0.81 & 0.00 & 0.36 & 0.00 & 0.20 & 0.00 & 0.15 & 0.00 & 0.15 & 0.34 & 6.60\\
 \hline
\end{tabular}}
\tablefoot{The maximal change in the guiding radii in our simulations for various parameters of the spiral structure. We investigated the change in time intervals of $1~\mbox{Gyr}$ (first six columns) and a longer interval of $4~\mbox{Gyr}$ (last column). The negative $\mbox{d}R_{g}$ represents an inward change in $R_g$, while the positive $\mbox{d}R_{g}$ stands for an outward change of $R_g$. The values represent the percentages of stars experiencing a maximal change in the guiding radius in the respective interval of d$R_g$. For each time bin, the left sub-column represents the DM configuration, while the right sub-column corresponds to the MOND configuration. The units of $\Omega_{sp}$ and $A_{sp}$ are $\mbox{km}~\mbox{s}^{-1}~\mbox{kpc}^{-1}$ and $\mbox{km}^{2}~\mbox{s}^{-2}~\mbox{kpc}^{-1}$, respectively.}
\end{table*}

We used the change in the guiding radius (Eq. \ref{R_gyr_Halle}) of stars to investigate and quantify changes in the radial structure of the galactic disc. We evaluated the simulation in time bins of $1~\mbox{Gyr}$. For each star, we calculated the maximum change in the guiding radius, $|\mbox{d}R_g| = R_{g_{max}} - R_{g_{min}}$, during each time period. If a star reached the distance $R_{g_{min}}$ earlier than the distance $R_{g_{max}}$, we set $\mbox{d}R_g$ as positive, which corresponds to outward changes in $R_g$; otherwise, we denoted it as negative, which represents inward changes in $R_g$. Our results are summarised in Table \ref{Tab_radmig}, using radial bins of $1$ kpc.

For the DM configuration, there is no significant trend in the time evolution of the maximal change in the guiding radius. However, time variations are present in the MOND configuration, where a larger fraction of stars tend to move inwards during the first half of the integration than during the second half. This means that a smaller portion of stars exhibit a small change in $R_g$ ($|\mbox{d}R_g| < 0.5~\mbox{kpc}$) in the first half of the simulation, compared to the last $3~\mbox{Gyr}$ of the simulation. In all five MOND configurations, the fraction of stars with small changes in $R_g$ increases with time. This effect might be partially caused by using the same set of initial conditions for the DM and MOND regimes, although the theoretical rotation curve is slightly lower for the MOND configuration than for the DM one. The fractions of stars with small changes in $R_g$ in the DM configurations span a wider range, from around $40\%$ up to $85\%$, depending on the parameters of the spiral structure. On the other hand, the MOND configurations show much lower fractions of stars with small changes in $R_g$, varying from $18\%$ to $45\%$ (depending on the spiral parameters).

Furthermore, for each configuration of the spiral parameters, the MOND simulation shows more significant changes in $R_g$, with a higher fraction of stars experiencing a larger change in the guiding radius, $|\mbox{d}R_g|$, for MOND compared to the DM approach. Two MOND configurations with $\Omega_{sp}=25~\mbox{km}~\mbox{s}^{-1}~\mbox{kpc}^{-1}$ and $A_{sp}= \{ 1000, 1200 \} ~\mbox{km}^2~\mbox{s}^{-2}~\mbox{kpc}^{-1}$ have more stars with a maximal change in the guiding radius of $0.5~\mbox{kpc} < |\mbox{d}R_{g}| < 1.5~\mbox{kpc}$ than stars with $|\mbox{d}R_g| < 0.5~\mbox{kpc}$. This effect is much more prominent when we consider smaller bins for radial distance (e.g. $200~\mbox{pc}$).

To capture the overall trend, we investigated $\mbox{d}R_{g}$ within a $4~\mbox{Gyr}$ time interval ($2-6~\mbox{Gyr}$, the last column in Table \ref{Tab_radmig}). A double peak distribution is present for all MOND configurations with $\Omega_{sp}=\{19, 25\}~\mbox{km}~\mbox{s}^{-1}~\mbox{kpc}^{-1}$ as well as the DM configuration with $\Omega_{sp}=25 \mbox{km}~\mbox{s}^{-1}~\mbox{kpc}^{-1}$ and $A_{sp}= 1200~\mbox{km}^2~\mbox{s}^{-2}~\mbox{kpc}^{-1}$, where the bin with $|\mbox{d}R_g| < 0.5~\mbox{kpc}$ contains fewer stars than the surrounding bins. The ratio, $\frac{|\mathrm{d}R_{g,\text{MOND}}| > 1.5~ \mathrm{kpc}}{|\mathrm{d}R_{g,\text{DM}}| > 1.5~\mathrm{kpc}}$, of the fraction of stars with a maximum change in the guiding radius larger than $1.5~\mbox{kpc}$ during the time interval $2-6~\mbox{Gyr}$ in the MOND and DM configurations, is presented in Table \ref{Tab_radmig2}. All MOND configurations show a higher portion of stars with a maximum change in the guiding radius larger than $1.5~\mbox{kpc}$ than the DM configurations. This ratio increases with the amplitude of the spiral pattern, $A_{sp}$, but decreases with the spiral pattern speed from around $5.4$ for $\Omega_{sp}=19~\mbox{km}~\mbox{s}^{-1}~\mbox{kpc}^{-1}$ to about $1.9$ for $\Omega_{sp}=31~\mbox{km}~\mbox{s}^{-1}~\mbox{kpc}^{-1}$. 

\begin{table}
\caption{\centering Ratio of the MOND and DM fractions of stars with significant changes of the guiding radius.}
\label{Tab_radmig2}
\centering
\begin{tabular}{|c|c|c|}
 \hline
 $\Omega_{sp}$ & $A_{sp}$ & $\frac{|\mathrm{d}R_{g,\text{MOND}}| > 1.5~ \mathrm{kpc}}{|\mathrm{d}R_{g,\text{DM}}| > 1.5~\mathrm{kpc}}$\\
 \hline
 \hline
 19 & 1000 & 5.45\\
 \hline
  & 850 & 2.56\\
 25 & 1000 & 3.94\\
  & 1200 & 4.10\\
 \hline
 31 & 1000 & 1.88\\
 \hline
\end{tabular}
\tablefoot{A maximum change in the guiding radius larger than $1.5~\mbox{kpc}$ during the time interval $2-6~\mbox{Gyr}$ was considered. The units of $\Omega_{sp}$ and $A_{sp}$ are $\mbox{km}~\mbox{s}^{-1}~\mbox{kpc}^{-1}$ and $\mbox{km}^{2}~\mbox{s}^{-2}~\mbox{kpc}^{-1}$, respectively.}
\end{table}

We did not truncate the spiral arms at the OLR, as is discussed in \cite{2016ARA&A..54..667S}. The region beyond the OLR (see Table \ref{TableMAXI}) contains a relatively small number of test particles due to the exponential decrease in density. Moreover, the spiral arms also exhibit an exponential decrease (see Eq. \ref{poteSpiral}); thus, the effect of the spiral structure in these remote regions is minor. Consequently, the spiral truncation does not change the results of our simulations from a qualitative point of view.

\section{Discussion}\label{Discussion}
Our results are in good agreement with recent works that explore radial migration with various datasets and simulations. For instance, \cite{Feltzing2020} investigated astrometric data from the Gaia DR2 cross-matched with APOGEE DR14 and explored the blurring and churning in the MW. In their work, radial migration of a star is based on whether its formation radius lies within the pericentre and apocentre of its current orbit. They found that around $50-60\%$ of stars have radially migrated, while about $35-45\%$ of stars have radially migrated and are currently on circular orbits with $L_z/L_c>0.95$ ($L_z$ is angular momentum of a star in the z direction in cylindrical coordinates and $L_c$ is angular momentum in the z direction that the star would have if it were on a circular orbit characterised by the same energy as the current orbit). Moreover, around $10\%$ of the stars have experienced radial migration with $L_z/L_c>0.99$. Compared with our results (see Table \ref{Tab_radmig}), depending on the spiral structure parameters, $\Omega_{sp}$ and $A_{sp}$, around $15-60\%$ and $55-82\%$ of stars have experienced $|\mbox{d}R| > 0.5~\mbox{kpc}$ in the DM and MOND simulations, respectively.

Moreover, \cite{Feltzing2020} quantified the fraction of stars that have not undergone radial migration. They found that around $5-7\%$ of MW's stars have an absolute value for the difference between the current radius and the formation radius smaller than $200~\mbox{pc}$. Repeating the same analysis as is presented in Sec. \ref{extrmni migranti}, but with smaller radial bins of $200~\mbox{pc}$, we obtain $5-12\%$ and $2-5.8\%$ of stars with a maximum radial migration below $200~\mbox{pc}$ for the DM and MOND configurations, respectively (using $1~\mbox{Gyr}$ time bins).

The time evolution of a galactic disc in a late-type galaxy was recently examined by \cite{Bautista2021}, where the authors simulated radial migration of stars in a steady potential. Their work is focussed on radial migration induced by the spiral structure (the galactic bar is omitted) with a potential that is the same as in our work (see Eq. \ref{poteSpiral}). The results presented in the form of the initial radius, $R_i$, versus the change in radius, $R_f-R_i$, for different time intervals (see their Fig. 3) show the same pattern with similar ridges (which correspond to different resonances) as in our work (Figs. \ref{Fig22.0}, \ref{Fig22}, and \ref{Fig22.2}). A similar pattern was observed by \cite{halle2018}. In agreement with our results, the most prominent ridges found by \cite{Bautista2021} are caused by the spiral arms' co-rotation resonance and spiral arms' ILR and OLR, while the higher-order spiral arm resonances are visible as well (with significantly smaller strength). They also investigated eccentricities, showing that the co-rotation and Lindblad resonances are responsible for complex radial migration. This is in agreement with our results (see the ridges corresponding to the co-rotation and Lindblad resonances in Figs. \ref{Fig11} and \ref{Fig11.2}). Furthermore, \cite{Bautista2021} investigated the influence of the spiral parameters $\Omega_{sp}$, $i$, and $r_d$ on radial migration, finding qualitatively similar results to ours, with the parameter $r_d$ being the most important.

The results from an N-body simulation presented by \citet{halle2018} suggest that the non-axisymmetrical features in MW-like galaxies develop naturally. Furthermore, their results show that the thin and thick disc components respond similarly to the non-axisymmetrical features, further justifying our approach with a planar galaxy. We have obtained similar results regarding the radial migration caused by co-rotation resonance with non-axisymmetrical components, in our case the spiral arms, where stars trapped at co-rotation resonance are churned outwards and inwards. We have also obtained consistent results about the nature of extreme migrators; namely, that their initial position must be close to the co-rotation resonance (see Figs. \ref{Fig22.0}, \ref{Fig22}, \ref{Fig22.2}, \ref{Fig11}, and \ref{Fig11.2}).

Resonance overlaps, such as those visible in Figs. \ref{freq_02N} and \ref{freq_02M}, were studied by \cite{2023MNRAS.523..991G} in the context of MW modelling. They found that resonance overlaps lead to an overestimation bias of the rotational velocity. In our model, this bias represents a limitation for dense regions close to the centre of our MW-like galaxy. Therefore, the significant changes in the galactocentric radius observed in our simulation in the inner region can be partially attributed to this bias. Together with stars being excluded from the simulation because they approach too close to the galactic centre, the accuracy of our simulations is limited in the inner regions of the galaxy.

\section{Conclusions}\label{Conclusions}

In this work, we have studied stellar migration in the galactic disc, considering the DM model and MOND theory to evaluate their effects on the evolution of the galaxy. We have explored the effects of the spiral arms amplitude, $A_{sp}$, and the pattern speed, $\Omega_{sp}$. Increasing both $A_{sp}$ and $\Omega_{sp}$ enhances the effect of resonances, leading to a more pronounced radial migration and changes in the galactocentric radius in both regimes. The strongest effect is generated by the co-rotation resonance with the spiral arms ($m=p=1$) and the OLRs with the galactic bar and spiral arms. We observe this behaviour in both the DM and MOND approaches.
Radial migration and changes in the galactocentric radius, driven by resonances with the galactic bar and spiral arms, are more pronounced in MOND. We observe more significant changes in the frequency phase space throughout the time evolution in the MOND case compared to the DM case. Compared to the DM approach, in the MOND case, we observe up to five times as many stars with a maximum change in the guiding radius of more than $1.5~\mbox{kpc}$ during the last $4~\mbox{Gyr}$ of the simulation. However, this effect becomes less pronounced with an increase in the spiral pattern speed. In future work, we plan to generalise the model by incorporating the vertical dimension and to study the resonances in the DM and the MOND regimes in 3D.

\begin{acknowledgements}
We thank the anonymous referee and Žofia Chrobáková for their helpful comments, which improved this paper. RN was funded by the EU NextGenerationEU through the Recovery and Resilience Plan for Slovakia under the project No. 09I03-03-V04-00137. FJ and MS were supported by the VEGA - the Slovak Grant Agency for Science, grant No. 1/0761/21. This work was supported in part through the Comenius University in Bratislava CLARA@UNIBA.SK high-performance computing facilities, services and staff expertise of Centre for Information Technology. https://uniba.sk/en/HPC-Clara.
\end{acknowledgements}

\bibliographystyle{aa} 
\bibliography{aa52112-24corr-v2.bbl}

\begin{thebibliography}{67}
\expandafter\ifx\csname natexlab\endcsname\relax\def\natexlab#1{#1}\fi

\bibitem[{{Anders} {et~al.}(2017){Anders}, {Chiappini}, {Minchev}, {Miglio}, {Montalb{\'a}n}, {Mosser}, {Rodrigues}, {Santiago}, {Baudin}, {Beers}, {da Costa}, {Garc{\'\i}a}, {Garc{\'\i}a-Hern{\'a}ndez}, {Holtzman}, {Maia}, {Majewski}, {Mathur}, {Noels-Grotsch}, {Pan}, {Schneider}, {Schultheis}, {Steinmetz}, {Valentini}, \& {Zamora}}]{Anders2017}
{Anders}, F., {Chiappini}, C., {Minchev}, I., {et~al.} 2017, \aap, 600, A70

\bibitem[{{Antoja} {et~al.}(2011){Antoja}, {Figueras}, {Romero-G{\'o}mez}, {Pichardo}, {Valenzuela}, \& {Moreno}}]{antoja2011}
{Antoja}, T., {Figueras}, F., {Romero-G{\'o}mez}, M., {et~al.} 2011, \mnras, 418, 1423

\bibitem[{{Bakos} {et~al.}(2008){Bakos}, {Trujillo}, \& {Pohlen}}]{Bakos2008}
{Bakos}, J., {Trujillo}, I., \& {Pohlen}, M. 2008, \apjl, 683, L103

\bibitem[{{Binney} \& {Tremaine}(2008)}]{binney-tremaine2008}
{Binney}, J. \& {Tremaine}, S. 2008, {Galactic Dynamics: Second Edition}

\bibitem[{{Bovy} {et~al.}(2019){Bovy}, {Leung}, {Hunt}, {Mackereth}, {Garc{\'\i}a-Hern{\'a}ndez}, \& {Roman-Lopes}}]{Bovy2019}
{Bovy}, J., {Leung}, H.~W., {Hunt}, J. A.~S., {et~al.} 2019, \mnras, 490, 4740

\bibitem[{{Brunetti} {et~al.}(2011){Brunetti}, {Chiappini}, \& {Pfenniger}}]{2011A&A...534A..75B}
{Brunetti}, M., {Chiappini}, C., \& {Pfenniger}, D. 2011, \aap, 534, A75

\bibitem[{{Ceverino} \& {Klypin}(2007)}]{Ceverino2007}
{Ceverino}, D. \& {Klypin}, A. 2007, \mnras, 379, 1155

\bibitem[{{Chen} \& {Zhao}(2020)}]{Chen20}
{Chen}, Y.~Q. \& {Zhao}, G. 2020, \mnras, 495, 2673

\bibitem[{{Contopoulos} \& {Grosbol}(1986)}]{contopoulos1986}
{Contopoulos}, G. \& {Grosbol}, P. 1986, \aap, 155, 11

\bibitem[{{Daniel} {et~al.}(2019){Daniel}, {Schaffner}, {McCluskey}, {Fiedler Kawaguchi}, \& {Loebman}}]{2019ApJ...882..111D}
{Daniel}, K.~J., {Schaffner}, D.~A., {McCluskey}, F., {Fiedler Kawaguchi}, C., \& {Loebman}, S. 2019, \apj, 882, 111

\bibitem[{{Daniel} \& {Wyse}(2015)}]{2015MNRAS.447.3576D}
{Daniel}, K.~J. \& {Wyse}, R. F.~G. 2015, \mnras, 447, 3576

\bibitem[{{Di Matteo} {et~al.}(2013){Di Matteo}, {Haywood}, {Combes}, {Semelin}, \& {Snaith}}]{2013A&A...553A.102D}
{Di Matteo}, P., {Haywood}, M., {Combes}, F., {Semelin}, B., \& {Snaith}, O.~N. 2013, \aap, 553, A102

\bibitem[{{Donlon} {et~al.}(2024){Donlon}, {Newberg}, {Sanderson}, {Bregou}, {Horta}, {Arora}, \& {Panithanpaisal}}]{Donlon2024}
{Donlon}, T., {Newberg}, H.~J., {Sanderson}, R., {et~al.} 2024, \mnras, 531, 1422

\bibitem[{{Eilers} {et~al.}(2019){Eilers}, {Hogg}, {Rix}, \& {Ness}}]{Eilers2019}
{Eilers}, A.-C., {Hogg}, D.~W., {Rix}, H.-W., \& {Ness}, M.~K. 2019, \apj, 871, 120

\bibitem[{{Feltzing} {et~al.}(2020){Feltzing}, {Bowers}, \& {Agertz}}]{Feltzing2020}
{Feltzing}, S., {Bowers}, J.~B., \& {Agertz}, O. 2020, \mnras, 493, 1419

\bibitem[{{Frankel} {et~al.}(2018){Frankel}, {Rix}, {Ting}, {Ness}, \& {Hogg}}]{Frankel2018}
{Frankel}, N., {Rix}, H.-W., {Ting}, Y.-S., {Ness}, M., \& {Hogg}, D.~W. 2018, \apj, 865, 96

\bibitem[{{Frankel} {et~al.}(2020){Frankel}, {Sanders}, {Ting}, \& {Rix}}]{Frankel20}
{Frankel}, N., {Sanders}, J., {Ting}, Y.-S., \& {Rix}, H.-W. 2020, \apj, 896, 15

\bibitem[{Freudenreich(1998)}]{freudenreich1998}
Freudenreich, H.~T. 1998, The Astrophysical Journal, 492, 495

\bibitem[{{Ghosh} {et~al.}(2023){Ghosh}, {Trick}, \& {Green}}]{2023MNRAS.523..991G}
{Ghosh}, S., {Trick}, W.~H., \& {Green}, G.~M. 2023, \mnras, 523, 991

\bibitem[{{Grand} {et~al.}(2016){Grand}, {Springel}, {G{\'o}mez}, {Marinacci}, {Pakmor}, {Campbell}, \& {Jenkins}}]{Grand2016b}
{Grand}, R. J.~J., {Springel}, V., {G{\'o}mez}, F.~A., {et~al.} 2016, \mnras, 459, 199

\bibitem[{{Halle} {et~al.}(2015){Halle}, {Di Matteo}, {Haywood}, \& {Combes}}]{2015A&A...578A..58H}
{Halle}, A., {Di Matteo}, P., {Haywood}, M., \& {Combes}, F. 2015, \aap, 578, A58

\bibitem[{{Halle} {et~al.}(2018){Halle}, {Di Matteo}, {Haywood}, \& {Combes}}]{halle2018}
{Halle}, A., {Di Matteo}, P., {Haywood}, M., \& {Combes}, F. 2018, \aap, 616, A86

\bibitem[{{Haywood}(2008)}]{Haywood_2008}
{Haywood}, M. 2008, \mnras, 388, 1175

\bibitem[{{Kla{\v{c}}ka}(2019)}]{Klacka2019}
{Kla{\v{c}}ka}, J. 2019, arXiv e-prints, arXiv:1904.04074

\bibitem[{{Kla{\v{c}}ka} {et~al.}(2012){Kla{\v{c}}ka}, {Nagy}, \& {Jur{\v{c}}i}}]{klacka2012}
{Kla{\v{c}}ka}, J., {Nagy}, R., \& {Jur{\v{c}}i}, M. 2012, \mnras, 427, 358

\bibitem[{{Kubryk} {et~al.}(2013){Kubryk}, {Prantzos}, \& {Athanassoula}}]{2013MNRAS.436.1479K}
{Kubryk}, M., {Prantzos}, N., \& {Athanassoula}, E. 2013, \mnras, 436, 1479

\bibitem[{{Kubryk} {et~al.}(2015){Kubryk}, {Prantzos}, \& {Athanassoula}}]{Kubryk2015}
{Kubryk}, M., {Prantzos}, N., \& {Athanassoula}, E. 2015, \aap, 580, A126

\bibitem[{{Lehmann} {et~al.}(2024){Lehmann}, {Feltzing}, {Feuillet}, \& {Kordopatis}}]{2024MNRAS.533..538L}
{Lehmann}, C., {Feltzing}, S., {Feuillet}, D., \& {Kordopatis}, G. 2024, \mnras, 533, 538

\bibitem[{{Lelli} {et~al.}(2016){Lelli}, {McGaugh}, \& {Schombert}}]{SPARCmaster}
{Lelli}, F., {McGaugh}, S.~S., \& {Schombert}, J.~M. 2016, \aj, 152, 157

\bibitem[{{Lian} {et~al.}(2022){Lian}, {Zasowski}, {Hasselquist}, {Holtzman}, {Boardman}, {Cunha}, {Fern{\'a}ndez-Trincado}, {Frinchaboy}, {Garcia-Hernandez}, {Nitschelm}, {Lane}, {Thomas}, \& {Zhang}}]{Lian22}
{Lian}, J., {Zasowski}, G., {Hasselquist}, S., {et~al.} 2022, \mnras, 511, 5639

\bibitem[{{Loebman} {et~al.}(2016){Loebman}, {Debattista}, {Nidever}, {Hayden}, {Holtzman}, {Clarke}, {Ro{\v{s}}kar}, \& {Valluri}}]{Loebmann2016}
{Loebman}, S.~R., {Debattista}, V.~P., {Nidever}, D.~L., {et~al.} 2016, \apjl, 818, L6

\bibitem[{{Loebman} {et~al.}(2011){Loebman}, {Ro{\v{s}}kar}, {Debattista}, {Ivezi{\'c}}, {Quinn}, \& {Wadsley}}]{Loebman2011}
{Loebman}, S.~R., {Ro{\v{s}}kar}, R., {Debattista}, V.~P., {et~al.} 2011, \apj, 737, 8

\bibitem[{{L{\'o}pez-Corredoira} \& {Sylos Labini}(2019)}]{2019A&A...621A..48L}
{L{\'o}pez-Corredoira}, M. \& {Sylos Labini}, F. 2019, \aap, 621, A48

\bibitem[{{Lu} {et~al.}(2024){Lu}, {Minchev}, {Buck}, {Khoperskov}, {Steinmetz}, {Libeskind}, {Cescutti}, {Freeman}, \& {Ratcliffe}}]{Lu2022}
{Lu}, Y.~L., {Minchev}, I., {Buck}, T., {et~al.} 2024, \mnras, 535, 392

\bibitem[{{Lucchini} {et~al.}(2024){Lucchini}, {D'Onghia}, \& {Aguerri}}]{Lucchini2024}
{Lucchini}, S., {D'Onghia}, E., \& {Aguerri}, J. A.~L. 2024, \mnras, 531, L14

\bibitem[{{Lépine} {et~al.}(2011){Lépine}, {Roman-Lopes}, {Abraham}, {Junquiera}, \& {Mishurov}}]{lepine2011a}
{Lépine}, J. R.~D., {Roman-Lopes}, A., {Abraham}, Z., {Junquiera}, T.~C., \& {Mishurov}, Y.~N. 2011, \mnras, 414, 1607

\bibitem[{{Mart{\'\i}nez-Barbosa} {et~al.}(2015){Mart{\'\i}nez-Barbosa}, {Brown}, \& {Portegies Zwart}}]{martinez-barbosa2015}
{Mart{\'\i}nez-Barbosa}, C.~A., {Brown}, A.~G.~A., \& {Portegies Zwart}, S. 2015, \mnras, 446, 823

\bibitem[{{Mart{\'\i}nez-Bautista} {et~al.}(2021){Mart{\'\i}nez-Bautista}, {Vel{\'a}zquez}, {P{\'e}rez-Villegas}, \& {Moreno}}]{Bautista2021}
{Mart{\'\i}nez-Bautista}, G., {Vel{\'a}zquez}, H., {P{\'e}rez-Villegas}, A., \& {Moreno}, E. 2021, \mnras, 504, 5919

\bibitem[{{McGaugh} {et~al.}(2016){McGaugh}, {Lelli}, \& {Schombert}}]{mcgaugh2016}
{McGaugh}, S.~S., {Lelli}, F., \& {Schombert}, J.~M. 2016, \prl, 117, 201101

\bibitem[{{McMillan}(2017)}]{mcmillan2017}
{McMillan}, P.~J. 2017, \mnras, 465, 76

\bibitem[{{Merrow} {et~al.}(2024){Merrow}, {Grand}, {Fragkoudi}, \& {Martig}}]{Merrow2024}
{Merrow}, A., {Grand}, R. J.~J., {Fragkoudi}, F., \& {Martig}, M. 2024, \mnras, 531, 1520

\bibitem[{{Minchev} {et~al.}(2014){Minchev}, {Chiappini}, \& {Martig}}]{Minchev2014}
{Minchev}, I., {Chiappini}, C., \& {Martig}, M. 2014, \aap, 572, A92

\bibitem[{{Minchev} \& {Famaey}(2010)}]{minchev2010}
{Minchev}, I. \& {Famaey}, B. 2010, \apj, 722, 112

\bibitem[{{Minchev} {et~al.}(2011){Minchev}, {Famaey}, {Combes}, {Di Matteo}, {Mouhcine}, \& {Wozniak}}]{Minchev2011}
{Minchev}, I., {Famaey}, B., {Combes}, F., {et~al.} 2011, \aap, 527, A147

\bibitem[{{Minchev} {et~al.}(2012){Minchev}, {Famaey}, {Quillen}, {Dehnen}, {Martig}, \& {Siebert}}]{Minchev2012}
{Minchev}, I., {Famaey}, B., {Quillen}, A.~C., {et~al.} 2012, \aap, 548, A127

\bibitem[{{Navarro} {et~al.}(1996){Navarro}, {Frenk}, \& {White}}]{nfw1996}
{Navarro}, J.~F., {Frenk}, C.~S., \& {White}, S. D.~M. 1996, \apj, 462, 563

\bibitem[{{Okalidis} {et~al.}(2022){Okalidis}, {Grand}, {Yates}, \& {Springel}}]{Okalidis2022}
{Okalidis}, P., {Grand}, R. J.~J., {Yates}, R.~M., \& {Springel}, V. 2022, \mnras, 514, 5085

\bibitem[{{Quillen} \& {Minchev}(2005)}]{quillen-minchev2005}
{Quillen}, A.~C. \& {Minchev}, I. 2005, \aj, 130, 576

\bibitem[{{Quillen} {et~al.}(2009){Quillen}, {Minchev}, {Bland-Hawthorn}, \& {Haywood}}]{2009MNRAS.397.1599Q}
{Quillen}, A.~C., {Minchev}, I., {Bland-Hawthorn}, J., \& {Haywood}, M. 2009, \mnras, 397, 1599

\bibitem[{{Roberts} {et~al.}(1979){Roberts}, {Huntley}, \& {van Albada}}]{roberts1979}
{Roberts}, W.~W., J., {Huntley}, J.~M., \& {van Albada}, G.~D. 1979, \apj, 233, 67

\bibitem[{{Ro{\v{s}}kar} {et~al.}(2013){Ro{\v{s}}kar}, {Debattista}, \& {Loebman}}]{Roskar2013}
{Ro{\v{s}}kar}, R., {Debattista}, V.~P., \& {Loebman}, S.~R. 2013, \mnras, 433, 976

\bibitem[{{Ro{\v{s}}kar} {et~al.}(2008{\natexlab{a}}){Ro{\v{s}}kar}, {Debattista}, {Quinn}, {Stinson}, \& {Wadsley}}]{Roskar2008b}
{Ro{\v{s}}kar}, R., {Debattista}, V.~P., {Quinn}, T.~R., {Stinson}, G.~S., \& {Wadsley}, J. 2008{\natexlab{a}}, \apjl, 684, L79

\bibitem[{{Ro{\v{s}}kar} {et~al.}(2008{\natexlab{b}}){Ro{\v{s}}kar}, {Debattista}, {Stinson}, {Quinn}, {Kaufmann}, \& {Wadsley}}]{Roskar2008}
{Ro{\v{s}}kar}, R., {Debattista}, V.~P., {Stinson}, G.~S., {et~al.} 2008{\natexlab{b}}, \apjl, 675, L65

\bibitem[{{Ruiz-Lara} {et~al.}(2017){Ruiz-Lara}, {P{\'e}rez}, {Florido}, {S{\'a}nchez-Bl{\'a}zquez}, {M{\'e}ndez-Abreu}, {S{\'a}nchez-Menguiano}, {S{\'a}nchez}, {Lyubenova}, {Falc{\'o}n-Barroso}, {van de Ven}, {Marino}, {de Lorenzo-C{\'a}ceres}, {Catal{\'a}n-Torrecilla}, {Costantin}, {Bland-Hawthorn}, {Galbany}, {Garc{\'\i}a-Benito}, {Husemann}, {Kehrig}, {M{\'a}rquez}, {Mast}, {Walcher}, {Zibetti}, {Ziegler}, \& {CALIFA Team}}]{Riuz-Lara2017}
{Ruiz-Lara}, T., {P{\'e}rez}, I., {Florido}, E., {et~al.} 2017, \aap, 604, A4

\bibitem[{{Sch{\"o}nrich} \& {Binney}(2009{\natexlab{a}})}]{Schoenrich2009a}
{Sch{\"o}nrich}, R. \& {Binney}, J. 2009{\natexlab{a}}, \mnras, 396, 203

\bibitem[{{Sch{\"o}nrich} \& {Binney}(2009{\natexlab{b}})}]{Schoenrich2009b}
{Sch{\"o}nrich}, R. \& {Binney}, J. 2009{\natexlab{b}}, \mnras, 399, 1145

\bibitem[{Sellwood \& Binney(2002)}]{sellwood2002radial}
Sellwood, J.~A. \& Binney, J. 2002, Monthly Notices of the Royal Astronomical Society, 336, 785

\bibitem[{{Shu}(2016)}]{2016ARA&A..54..667S}
{Shu}, F.~H. 2016, \araa, 54, 667

\bibitem[{{Solway} {et~al.}(2012){Solway}, {Sellwood}, \& {Sch{\"o}nrich}}]{Solway2012}
{Solway}, M., {Sellwood}, J.~A., \& {Sch{\"o}nrich}, R. 2012, \mnras, 422, 1363

\bibitem[{{Trick}(2022)}]{Trick2022}
{Trick}, W.~H. 2022, \mnras, 509, 844

\bibitem[{{Trick} {et~al.}(2021){Trick}, {Fragkoudi}, {Hunt}, {Mackereth}, \& {White}}]{Trick2021}
{Trick}, W.~H., {Fragkoudi}, F., {Hunt}, J. A.~S., {Mackereth}, J.~T., \& {White}, S. D.~M. 2021, \mnras, 500, 2645

\bibitem[{{Vera-Ciro} {et~al.}(2014){Vera-Ciro}, {D'Onghia}, {Navarro}, \& {Abadi}}]{vera-ciro2014}
{Vera-Ciro}, C., {D'Onghia}, E., {Navarro}, J., \& {Abadi}, M. 2014, \apj, 794, 173

\bibitem[{Virtanen {et~al.}(2020)Virtanen, Gommers, Oliphant, Haberland, Reddy, Cournapeau, Burovski, Peterson, Weckesser, Bright, {van der Walt}, Brett, Wilson, Millman, Mayorov, Nelson, Jones, Kern, Larson, Carey, Polat, Feng, Moore, {VanderPlas}, Laxalde, Perktold, Cimrman, Henriksen, Quintero, Harris, Archibald, Ribeiro, Pedregosa, {van Mulbregt}, \& {SciPy 1.0 Contributors}}]{2020SciPy}
Virtanen, P., Gommers, R., Oliphant, T.~E., {et~al.} 2020, Nature Methods, 17, 261

\bibitem[{{Viscasillas V{\'a}zquez} {et~al.}(2023){Viscasillas V{\'a}zquez}, {Magrini}, {Spina}, {Tautvai{\v{s}}ien{\.{e}}}, {Van der Swaelmen}, {Randich}, \& {Sacco}}]{2023A&A...679A.122V}
{Viscasillas V{\'a}zquez}, C., {Magrini}, L., {Spina}, L., {et~al.} 2023, \aap, 679, A122

\bibitem[{{Wang} \& {Zhao}(2013)}]{Wang2013}
{Wang}, Y. \& {Zhao}, G. 2013, \apj, 769, 4

\bibitem[{{Yoachim} {et~al.}(2012){Yoachim}, {Ro{\v{s}}kar}, \& {Debattista}}]{Yoachim2012}
{Yoachim}, P., {Ro{\v{s}}kar}, R., \& {Debattista}, V.~P. 2012, \apj, 752, 97

\bibitem[{{Zhang} {et~al.}(2021){Zhang}, {Chen}, \& {Zhao}}]{Zhang2021}
{Zhang}, H., {Chen}, Y., \& {Zhao}, G. 2021, \apj, 919, 52

\end{thebibliography}

\begin{appendix}
\onecolumn

\section{Distribution of changes in galactocentric radius for the rest of simulation runs}
\label{Appendix_0}

Here we present the continuation of Fig. \ref{Fig22.0}, showing the distribution of changes in galactocentric radius $R_{f} - R_{i}$ with respect to $R_{i}$ as in Sec. \ref{51radmig} for the rest of simulation runs (Figs. \ref{Fig22}, \ref{Fig22.2}).

\begin{figure*}[h!]
\centering 

\subfloat{%
  \includegraphics[width=1\textwidth]{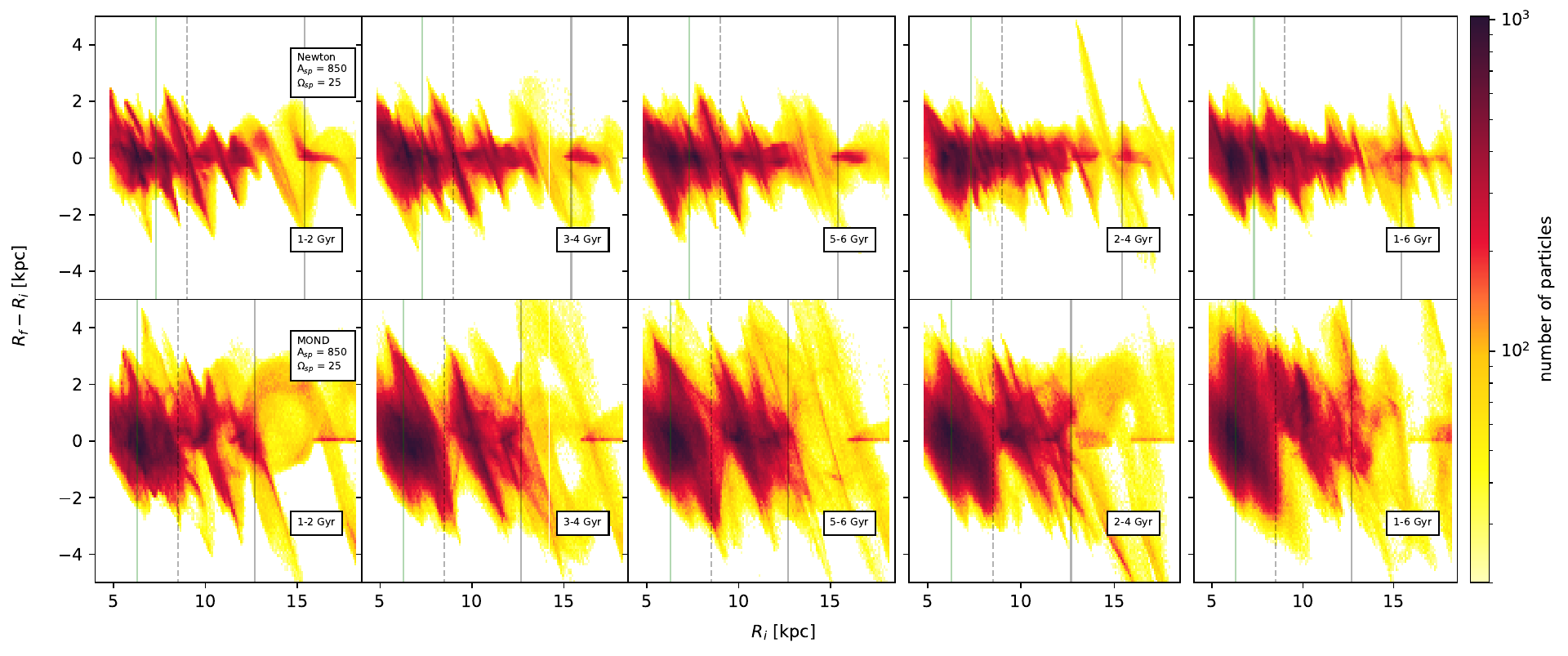}%
}

\subfloat{%
  \includegraphics[width=1\textwidth]{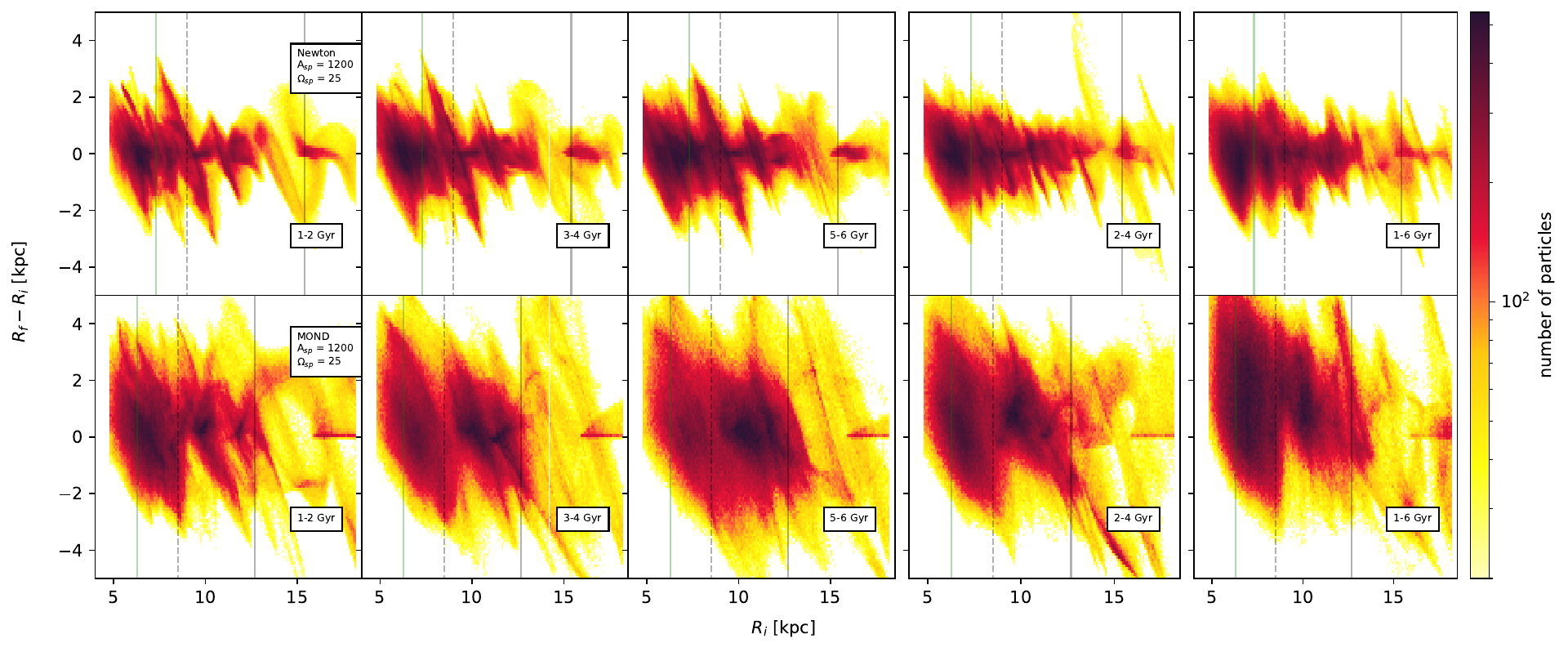}%
}
\caption{Distribution of changes in galactocentric radius $R_{f} - R_{i}$ as a function of $R_{i}$ for different time intervals during the $6~\mbox{Gyr}$ integration. Each set of figures represents different simulation run with parameters as listed in Tables \ref{parameters} and \ref{parameters2}. Vertical lines represent theoretical values of resonances: green lines represent resonances produced by the galactic bar, and grey lines represent resonances caused by the spiral arms. For both cases, the solid lines represent OLRs and the dashed line the $m=p=1$ co-rotation resonance.}
\label{Fig22}
\end{figure*}
\begin{figure*}[h!]
\centering 

\subfloat{%
  \includegraphics[width=1\textwidth]{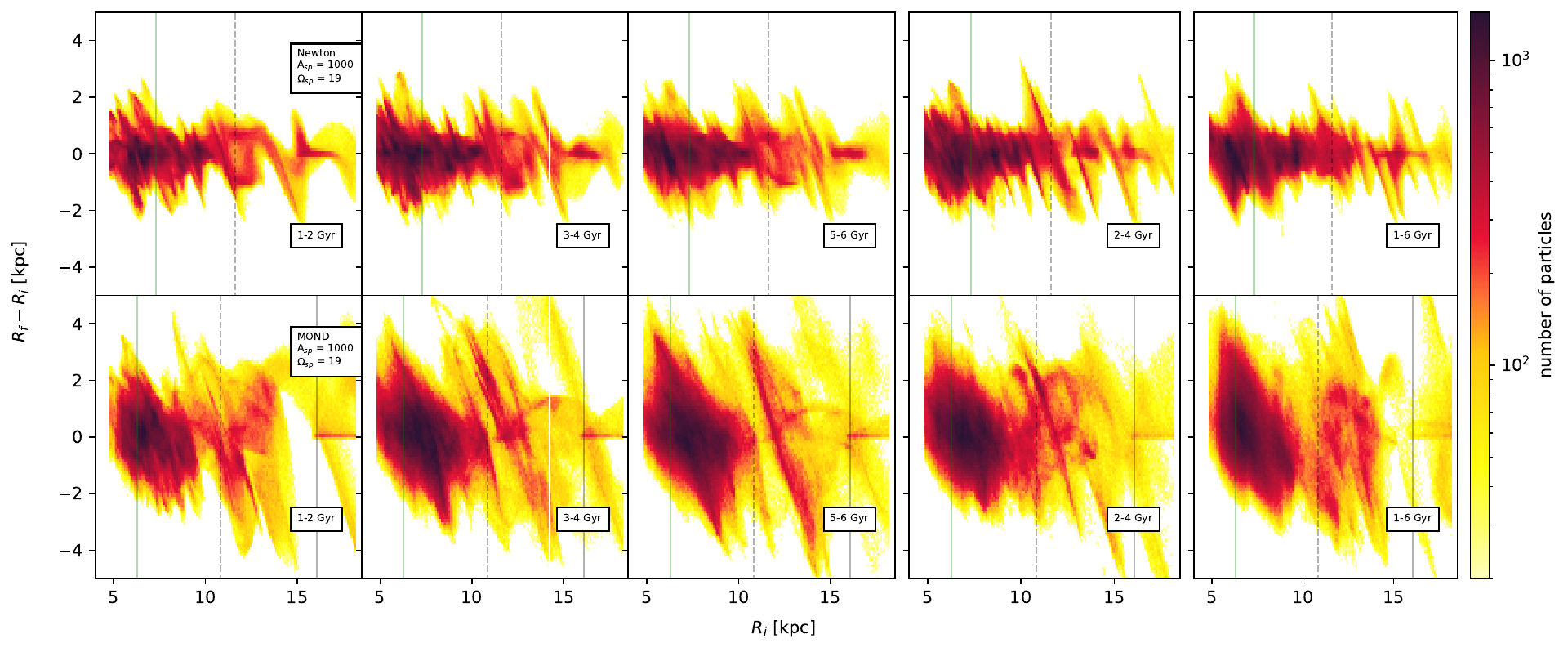}%
}

\subfloat{%
  \includegraphics[width=1\textwidth]{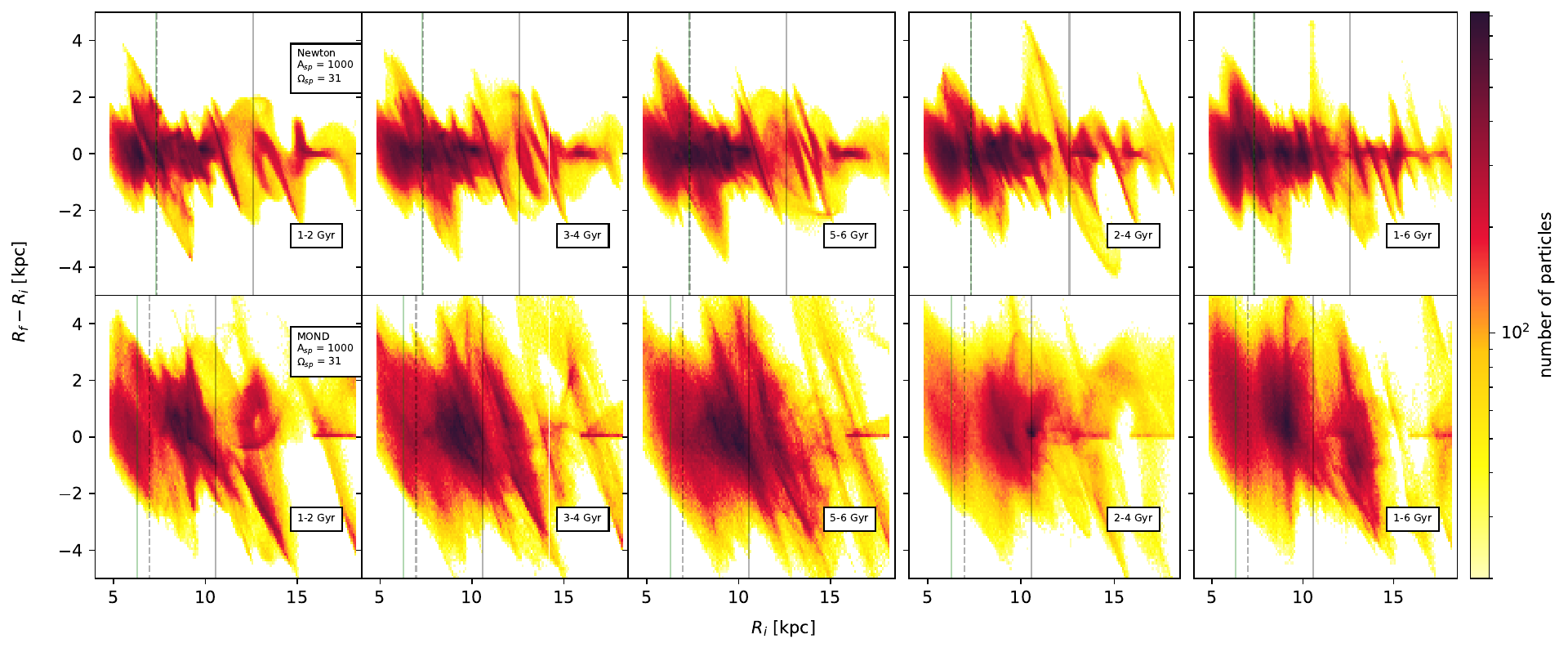}%
}

\caption{Continuation of Fig. \ref{Fig22}}
\label{Fig22.2}
\end{figure*}

\newpage

\section{Frequency phase space for $\Omega_{sp} = 19; 31~\mbox{km}~\mbox{s}^{-1}~\mbox{kpc}^{-1}$}
\label{Appendix_A}

\begin{figure}[h!]
\centering
\subfloat{\includegraphics[width=1\textwidth]{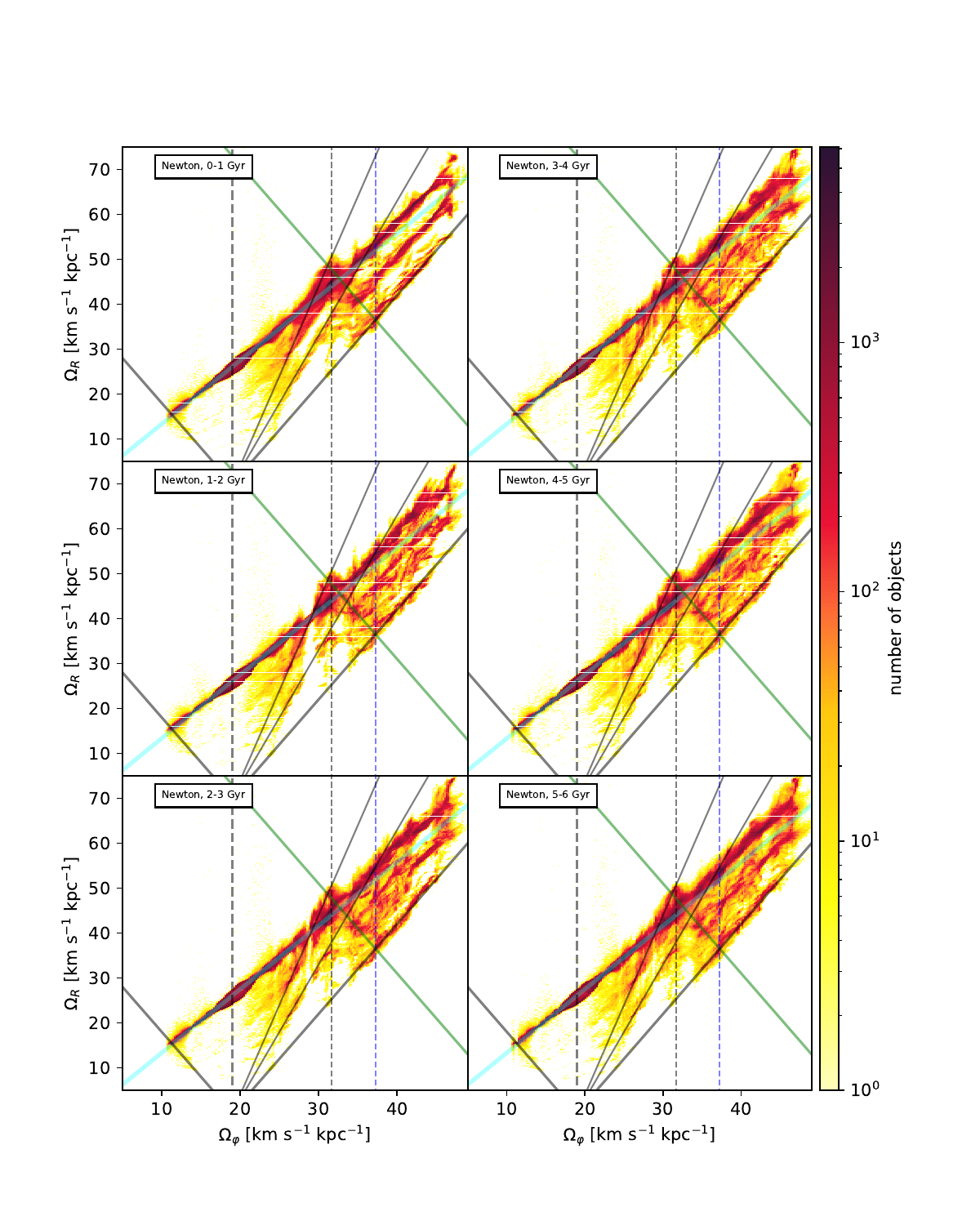}}
\caption{Time evolution of the frequency phase space with a time step of $1~\mbox{Gyr}$ for the simulation run DM\_19\_1000. The light blue curve indicates the epicycle approximation. Displayed resonances include spiral arms co-rotation resonances $m=p=1$ and $m=3$, $p=5$ (thick and thin grey dashed lines respectively), co-rotation resonance $m=2$, $p=1$ with the superposition of the spiral arms and the galactic bar (blue dashed line), ILRs and OLRs with the spiral arms (thick solid grey lines), OLR with the galactic bar (thick green solid line), spiral arms resonances $l=-1$, $m=p=4$ and $l=-1$, $m=p=3$ (thin grey solid lines).}
\label{freq_01N}
\end{figure}

\begin{figure*}
\centering
\subfloat{\includegraphics[width=1\textwidth]{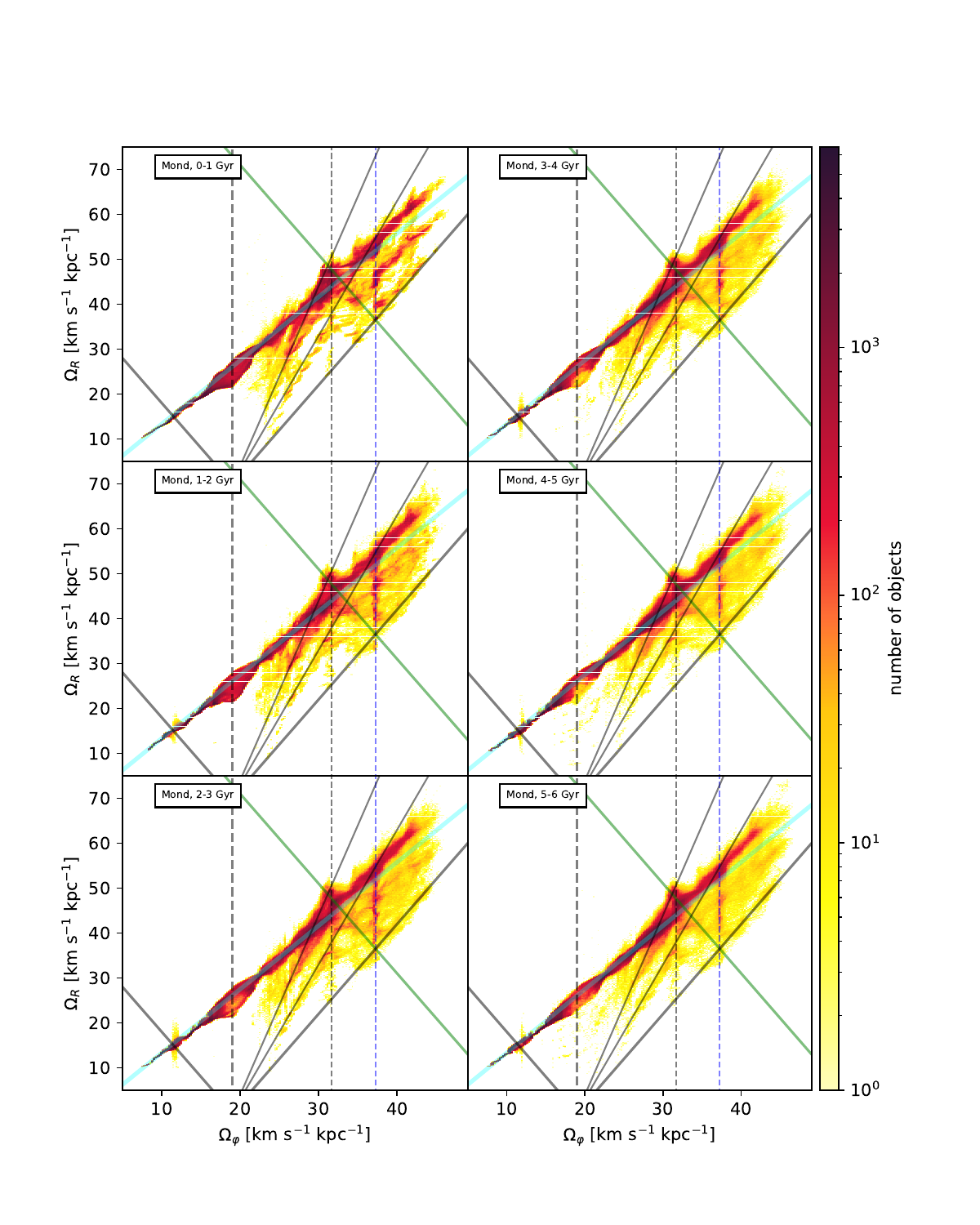}}
\caption{Same as Fig. \ref{freq_01N}, but for the simulation run MOND\_19\_1000.}
\label{freq_01M}
\end{figure*}

\begin{figure*}
\centering
\subfloat{\includegraphics[width=1\textwidth]{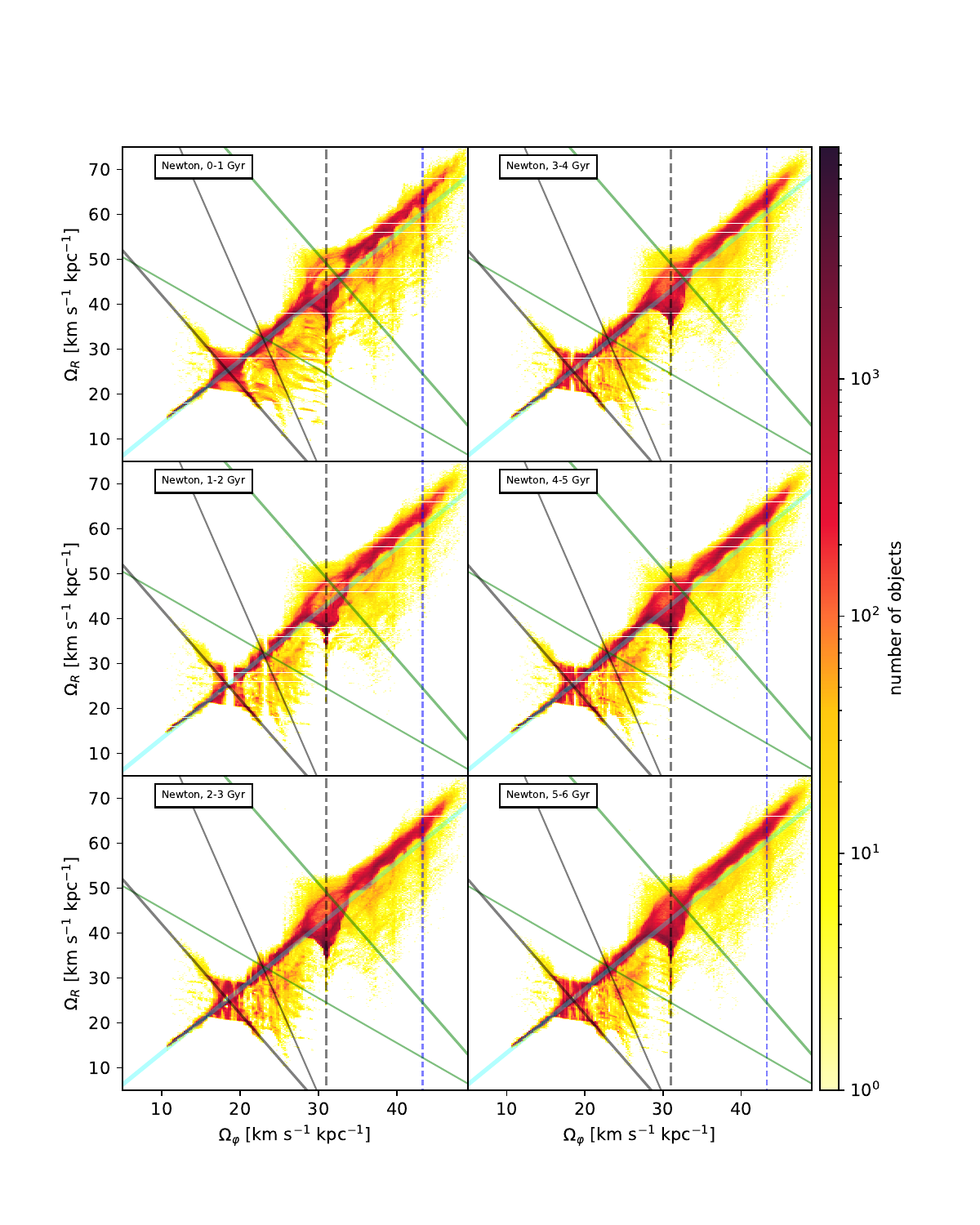}}
\caption{Same as Fig. \ref{freq_01N}, but for the simulation run DM\_31\_1000.}
\label{freq_03N}
\end{figure*}

\begin{figure*}
\centering
\subfloat{\includegraphics[width=1\textwidth]{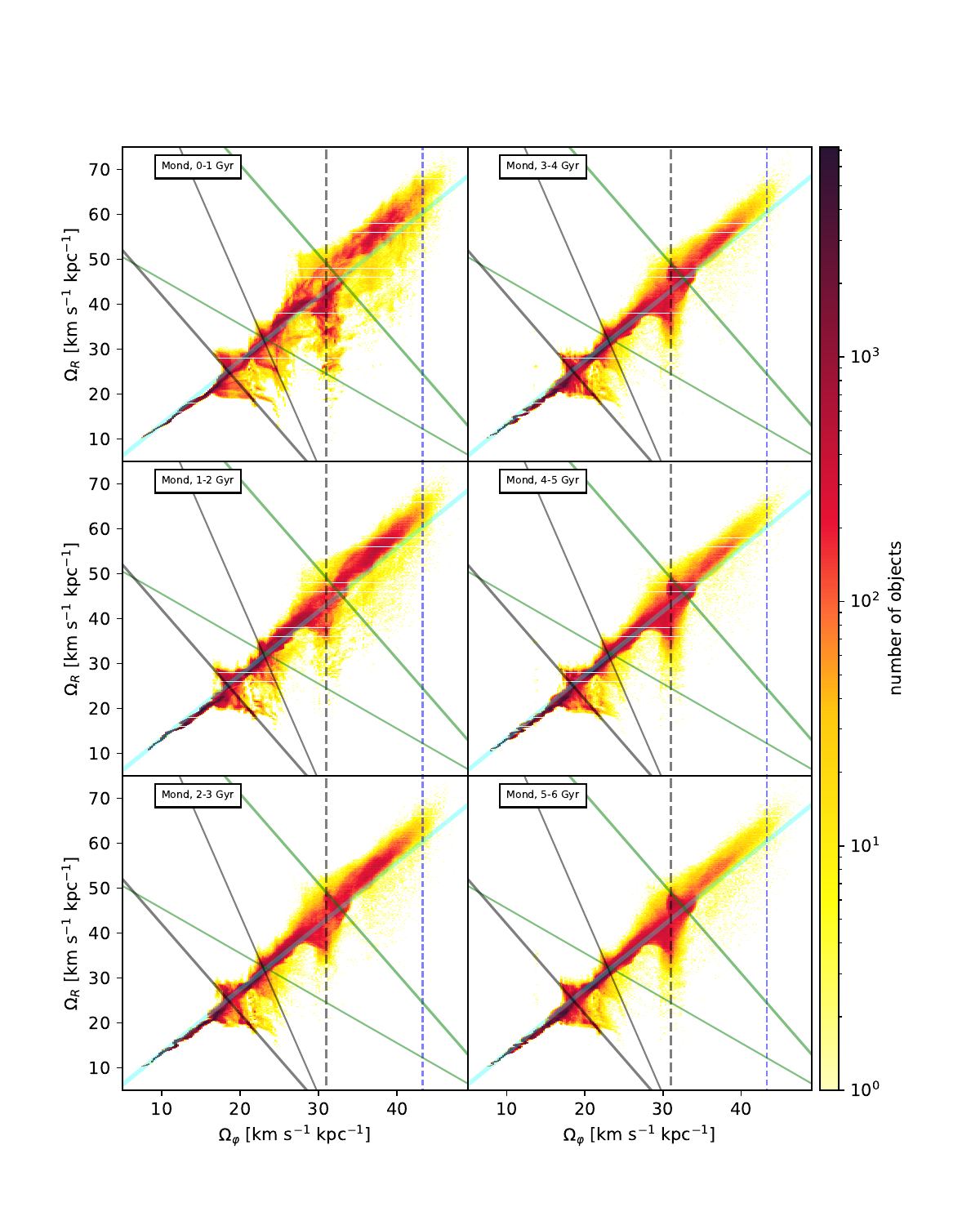}}
\caption{Same as Fig. \ref{freq_01N}, but for the simulation run MOND\_31\_1000.}
\label{freq_03M}
\end{figure*}

Here we present frequency maps as in Sec. \ref{reson} for simulation runs DM\_19\_1000 (Fig. \ref{freq_01N}), MOND\_19\_1000 (Fig. \ref{freq_01M}), DM\_31\_1000 (Fig. \ref{freq_03N}) and MOND\_31\_1000 (Fig. \ref{freq_03M}).

\end{appendix}

\end{document}